\def\PP{Yes} 

\def\yes{Yes} \def\no{No}
\newcommand{\IfPP}[1]{\ifx\PP\yes #1 \fi}
\newcommand{\IfMS}[1]{\ifx\PP\no #1 \fi}
\newcommand{\IfPPelse}[2]{\ifx\PP\yes #1 \else #2 \fi}
\IfPPelse{
  \documentclass{article}
  \usepackage{emulateapj}
  \usepackage{graphicx}
  \submitted{Accepted by The Astronomical Journal 29 July 1999 (in
    press November 1999)}
  }{
  \documentstyle[12pt,aasms4]{article}
  }

\newcommand{\Ha}{H$\alpha$}
\newcommand{\Hb}{H$\beta$}
\newcommand{\SII}{[\ion{S}{2}]\ 6731\AA}
\newcommand{\NII}{[\ion{N}{2}]\ 6583\AA}
\newcommand{\OI}{[\ion{O}{1}]\ 6300\AA}
\newcommand{\HeI}{\ion{He}{1}\ 5676\AA}
\newcommand{\OIII}{[\ion{O}{3}]\ 4959\AA}
\newcommand{\SIII}{[\ion{S}{3}]\ 6312\AA}
\newcommand{\SIIshort}{[\ion{S}{2}]}
\newcommand{\NIIshort}{[\ion{N}{2}]}
\newcommand{\OIshort}{[\ion{O}{1}]}

\newcommand{\OIIIshort}{[\ion{O}{3}]}
\newcommand{\SIIIshort}{[\ion{S}{3}]}
\IfPPelse{ 
  \newcommand{\SIIshortp}{[S~{\scriptsize II}]}
  \newcommand{\NIIshortp}{[N~{\scriptsize II}]}

  \newcommand{\NIIp}{[N~{\scriptsize II}]\ 6583\AA}
  
  \newcommand{\HeIp}{He~{\scriptsize I}\ 5676\AA}
  \newcommand{\OIIIp}{[O~{\scriptsize III}]\ 4959\AA}
  \newcommand{\SIIIp}{[S~{\scriptsize III}]\ 6312\AA}
  }{
  \newcommand{\SIIshortp}{[\protect\ion{S}{2}]}
  \newcommand{\NIIshortp}{[\protect\ion{N}{2}]}

  \newcommand{\NIIp}{[\protect\ion{N}{2}]\ 6583\AA}
  
  \newcommand{\HeIp}{\protect\ion{He}{1}\ 5676\AA}
  \newcommand{\OIIIp}{[\protect\ion{O}{3}]\ 4959\AA}
  \newcommand{\SIIIp}{[\protect\ion{S}{3}]\ 6312\AA}
  }
\newcommand{\etal}{et~al.}
\newcommand{\hii}{\mbox{\ion{H}{2}}}
\newcommand{\lya}{\mbox{Ly$\alpha$}}

\newcommand{\kms}{\mbox{$\,$km s$^{-1}$}}
\newcommand{\thC}{$\theta^1\,$C~Ori}
\newcommand{\thA}{$\theta^2\,$A~Ori}
\newcommand{\subsun}{M$_{\hbox{$\odot$}}$}
\newcommand{\peryr}{yr$^{-1}$}
\newcommand{\HST}{{\em HST\/}}
\newcommand{\Av}{A_V}
\newcommand{\tevap}{t_{\rm e}} 
\newcommand{\tdyn}{t_{\rm d}}
\newcommand{\rcrit}{r_{\rm cr}}
\newcommand{\smy}{\mbox{M}_\odot \, \mbox{yr}^{-1}}

\lefthead{HENNEY \& O'DELL}
\righthead{KECK SPECTROSCOPY OF THE ORION PROPLYDS}

\begin{document}

\title{A Keck High Resolution Spectroscopic Study of the Orion Nebula Proplyds}

\author{W. J. Henney} 

\affil{Instituto de Astronom\'{\i}a, Universidad Nacional Aut\'onoma
  de M\'exico, J. J. Tablada 1006, Lomas de Santa Mar\'{\i}a, 58090
  Morelia, Michoac\'an, M\'exico; will@astrosmo.unam.mx}

\authoremail{will@astrosmo.unam.mx}

\and

\author{C. R. O'Dell} 

\affil{Department of Space Physics and Astronomy, MS-108, Rice
  University, P.O. Box 1892, Houston, TX 77251; cro@rice.edu}

\authoremail{cro@rice.edu}

\begin{abstract}
  We present the results of spectroscopy of four bright proplyds in
  the Orion Nebula obtained at a velocity resolution of 6 \kms.  After
  careful isolation of the proplyd spectra from the confusing nebular
  radiation, the emission line profiles are compared with those
  predicted by realistic dynamic/photoionization models of the
  objects.  The spectral line widths show a clear correlation with
  ionization potential, which is consistent with the free expansion of
  a transonic, ionization-stratified, photoevaporating flow.  Fitting
  models of such a flow simultaneously to our spectra and \HST\ 
  emission line imaging provides direct measurements of the proplyd
  size, ionized density and outflow velocity. These measurements
  confirm that the ionization front in the proplyds is approximately
  D-critical and provide the most accurate and robust estimate to date
  of the proplyd mass loss rate.  Values of \(0.7\mbox{--}1.5 \times
  10^{-6} \smy\) are found for our spectroscopic sample, although
  extrapolating our results to a larger sample of proplyds implies
  that \(0.4\times 10^{-6} \smy\) is more typical of the proplyds as a
  whole.  In view of the reported limits on the masses of the
  circumstellar disks within the proplyds, the length of time that
  they can have been exposed to ionizing radiation should not greatly
  exceed \(10^4\) years --- a factor of 30 less than the mean age of
  the proplyd stars.  We review the various mechanisms that have been
  proposed to explain this situation, and conclude that none can
  plausibly work unless the disk masses are revised upwards by a
  substantial amount.
\end{abstract}


\section{Introduction}
\label{sec:Introduction}
Thick disks of circumstellar material seem to be an integral and
perhaps necessary component of newly formed stars.  When these stars
are found in or near an \hii\ region the possibilities of their
detection and survival become quite different than in regions lacking
massive early type stars.  The identification of this special class of
objects, called the proplyds (O'Dell \etal\ 1993), was first made in
the Orion Nebula, with two more distant objects having been found
(Stapelfeldt \etal\ 1997, Stecklum \etal\ 1998).  The Orion proplyds
are bright and have been detected as unresolved stars for many
decades; however, their special nature did not begin to become
unveiled until the emission line images of Laques \&\ Vidal (1979)
showed them to be compact ionized sources.  These observations were
extended with VLA observations at an angular resolution of 0.1\arcsec\ 
(Garay \etal\ 1987, Churchwell \etal\ 1987), with the latter paper
identifying their correct nature (young stars with circumstellar
clouds photoionized from the outside by \thC) as one of the possible
models, a view shared by Meaburn (1988) soon after.  However, it was
the first \HST\ WFPC2 images in the core of the Orion nebula that
clearly revealed the correctness of this interpretation (O'Dell \etal\ 
1993) and showed that the proplyds came in a variety of forms and
brightnesses.  What one sees depends very much on the location and
orientation of the proplyd.  Those distant from \thC\ and shielded
from its ionizing radiation appear dark in silhouette against the
bright nebular background, while those close to \thC\ have their outer
parts ionized by that star. This is revealed by bright cusps on the
side facing \thC, the surface brightness of which scales with distance
from the ionizing star in the manner expected (O'Dell 1998).  In several
of the bright proplyds one can see the inner neutral dust disk in
silhouette.  In all cases one can see a low mass star at the center of
the proplyd, except for two objects whose disk plane lies almost along
the line of sight(McCaughrean \& Stauffer 1994, O'Dell \&\ Wong 1996,
McCaughrean \&\ O'Dell 1996, Chen \etal\ 1998).

A neutral hydrogen column density of about 10$^{17}\,$cm$^{-2}$ is all
that is required to reach an optical depth of unity in the Lyman
continuum radiation and thus an ionization front (IF). It is not
surprising, therefore, that the bright proplyds are bounded on the
side facing the ionizing star by a well defined ionization boundary.
That the bright rims are indeed IFs is established not only by the
variation of their surface brightness as the inverse square of the
distance from \thC, but also from the fact that this feature is also
bright in \OIshort\ and \SIIshort\ lines, telltale emission that can
only arise in an IF.  The most promising explanation for the observed
structure of the bright proplyds, is that a dense, slow neutral wind
is driven from the (largely molecular) circumstellar accretion disk by
the heating effect of non-ionizing FUV radiation.  This explanation
was initially hinted at by McCullough (1995), but was first developed
in depth by Johnstone \etal\ (1998). The flux of ionizing EUV photons
is not sufficient to ionize this wind all the way to its base and so
an IF forms that is offset from the disk surface by a few disk radii.
At present quite sophisticated models for the proplyds have been
generated within this framework (Henney \&\ Arthur 1998, Johnstone
\etal\ 1998, St\"orzer \&\ Hollenbach 1998,1999), all sharing the
features that neutral hydrogen is fed from the disk (where hydrogen
exists as H$_{2}$, Chen \etal\ 1998) into an extended outer atmosphere
whose side facing \thC\ or \thA\ is photoionized.  The St\"orzer \&\ 
Hollenbach (1998,1999) models contain the most detailed treatment of
the photodissociated neutral flow and establish that they can explain
both the intensity of the H$_{2}$ at the boundary of the inner disk
and also the \OIshort\ emission that has been observed there (Bally
\etal\ 1998b) and cannot arise from collisional excitation by
photoelectrons.  Models in which the accretion disk is directly
ionized by EUV radiation (Henney \etal\ 1996, Richling \&\ Yorke 1999)
may apply to the smallest proplyds, which are closest to \thC.

The fact that material should freely flow away from the ionization
front (Dyson 1968), hence producing mass loss from the proplyds, was
recognized as early as the Churchwell \etal\ (1987) discovery paper,
with current models predicting rates of about 10$^{-7}$--10$^{-6}$
\subsun\ \peryr.  The mass loss rate depends on the IF radius,
\(r_0\), photoevaporating flow velocity, \(u_0\), and peak ionized
density, \(n_0\), as \( (\dot{M}/10^{-7}\smy) \simeq 2\mbox{--}4\,
(r_0/10^{15}\,\mbox{cm})^2 (u_0/13\,\mbox{km}\,\mbox{s}^{-1})
(n_0/10^6\,\mbox{cm}^{-3})\). The constant of proportionality in this
equation varies according to the relative importance of mass loss from
the sides and tails of the proplyds.  The masses of the disks can be
estimated from thermal emission from their dust component, with the
result that the most massive contain about 10$^{-2}$ \subsun (Mundy
\etal\ 1995, Lada \etal\ 1996, Bally \etal\ 1998b).  Combining these
masses with the theoretically predicted mass loss rates, indicates
disk destruction times of about a few times 10$^{4}$ to 10$^{5}$
years.  Although one knows that the Trapezium Cluster (TC) is quite
young (Hillenbrand 1997), the position of the stars on the luminosity
versus temperature diagram indicating an age of 300,000 to 10$^{6}$
years, the uncertainties of the pre-main sequence stellar models being
used prevents determination of an exact age or identification of an
extended period of star formation.  In any event, the theoretically
predicted disk survival ages are short compared with the age of the
cluster, and, furthermore, there is no depletion of proplyds as one
samples regions closer to \thC\ (O'Dell \&\ Wong 1996, Henney \&
Arthur 1998).  This means that we are missing some ingredient in the
modeling.  St\"orzer \&\ Hollenbach (1999) argue that the conundrum
can be resolved by assuming that the proplyds are in highly elliptical
orbits, thus spending only a small fraction of their time in the
vicinity of \thC.  O'Dell (1998) has argued that \lya\ photons trapped
within the confines of the nebula can produce a sufficent net inward
force to stop the proplyd loss of mass, but it should be noted that
what he derives as the amount of this constraining pressure would
actually be an upper limit.  Although there was some hope that one
could determine from the \HST\ images whether or not the ionized
portions of the proplyds were in static equilibrium, these aspirations
were dashed when Henney \&\ Arthur (1998) showed that the nearly
exponential surface brightness distributions within the bright cusps
could equally well be explained by both a freely expanding and a
static atmosphere.  This means that one must look to more direct means
of determining the mass loss rates.

The most powerful means available is to directly measure the velocity
of the flow of material off of the IF of individual proplyds.  This
approach presents particular observational challenges in obtaining and
analyzing the data and demands good supporting theoretical work.  The
observational challenges are primarily ones of accurately subtracting
the high velocity emission from any jets associated with the proplyds
and the low velocity background emission from the Orion nebula.
Supersonic flows from the proplyds were first detected by Meaburn
(1988) in the \OIIIshort~5007\AA\ line, who then characterized them
more fully with additional spectra (Meaburn \etal\ 1993, Massey \&\ 
Meaburn 1995, Henney \etal\ 1997).  These flows were studied in
additional lines by Hu (1996) and the entire nebula was completely
covered with Fabry-Perot spectra in \OIIIshort\ and \SIIshort\ 
(O'Dell, \etal\ 1997).  Sources with known jet flows are to be avoided
when looking for the photoevaporative flows, but one does not always
know beforehand that they are there. Fortunately the flows are
sufficiently large to be separable.  Accurate isolation of the proplyd
spectrum from the signal that contains both the proplyd and nebular
background emission is a greater challenge.  Both emissions come from
the fluoresence of Lyman continuum photons, with the closer proplyds
seeing a higher flux of these photons than the nebular IF, but the
proplyds are smaller than the typical seeing disk of a groundbased
telescope, so that their signal is often a small fraction of the total
signal.  Henney \etal\ (1997) made a thorough analysis of the best
proplyd spectra from the Massey \&\ Meaburn (1995) data set and
demonstrated that good isolation of velocity components near but not
at the systemic velocity could be done.  Even if good observational
isolation of the spectrum of the proplyd can be obtained, its correct
interpretation requires independent knowledge of as many of the
proplyd parameters as possible (e.~g., ionized density, IF radius,
orientation), for which the best images must be employed.  Once the
spectra are in hand and these parameters determined, then one must
have the predicted line profiles that would be expected for the static
or freely expanding ionized atmospheres.  It is these factors that
have guided the formulation of the study reported in this paper.

\section{Observations and Data Reduction}
\label{sec:Observations}

The exact determination of the mass loss rates of the proplyds has
repercussions that go well beyond whether or not the TC circumstellar
disks will survive.  The standard paradigm for star formation is that
the TC is more characteristic of where the bulk of stars form than
less rich clusters such as those found in the Taurus cloud where high
mass stars are not found.  If the TC is characteristic and the
circumstellar disks are rapidly destroyed, it seems unlikely that
planet formation will be ubiquitous.  It was, therefore, considered
worth the time and effort to pursue the question of the proplyd mass
loss rate with the most powerful groundbased telescope: the Keck I
observatory at Mauna Kea, Hawaii.  Fortunately, this opinion was
shared by the persons allocating Keck time within the NASA ``Origins
Program" and the observations were carried out as part of a joint
program with John Stauffer, who was studying low mass stars in the
same region with the same instrumental configuration.

The observations were made with the Keck I telescope the nights (UT)
of 5 and 6 December 1997 using the HIRES spectrograph and a
1024$\times$2048 Tektronix CCD.  The entrance slit was
0.862$\times$14\arcsec, projecting onto the detector with a velocity
resolution of 6.2$\pm$0.4 \kms\ and a scale of
0.382\arcsec/double-binned-pixel perpendicular to the dispersion.  The
HIRES spectrograph is a cross dispersed echelle spectrograph system
used in a configuration such that when the 14\arcsec\ long slit is
used, the ends of the highest orders employed (containing the \OIII\ 
and the \Hb\ 4861\AA\ lines) overlapped.  Fortunately, the Tektronix
CCD has a large linear range and the limiting signal imposed by the
digitizing system (2$^{16}$=65,536 counts) fell well beneath it.  This
large dynamic range allowed both strong and weak emission lines to be
recorded at a single echelle setting.  Wavelength calibration was
provided from exposures of a Thorium+Argon standard lamp.
Unfortunately, the Tektronix CCD had several pixel defects, whose
impact was minimized by adjusting the echelle and cross dispersing
grating orientations, but one such defect precluded using the images
of the stronger \OIIIshort\ 5007\AA\ line, which was not a serious
problem as the \OIII\ line from the same upper state always provided
an adequate signal.

The observing strategy was to select four proplyds which are located
in relatively smooth portions of the Orion nebula, are free from
adjacent bright stars, and represent a variety of sizes and
morphologies.  The objects observed were 170--337 (HST2), 177--431
(HST1), 182--413 (HST10), and 244--440. The coordinate based
designation system of O'Dell \&\ Wen (1994) will be used throughout
this paper, the previous parenthetical designations being those of the
serial listing in an earlier paper (O'Dell \etal\ 
1993).\footnote{244--440 was discovered later (O'Dell \&\ Wong 1996),
  but incorrectly included in their table of stellar objects because
  its emission line shell was much larger than in any other Orion
  proplyd.}  The first three objects have sizes (measured as the
distance between their bright cusp tips) of 0.4\arcsec, 0.6\arcsec,
and 1.2\arcsec\ (O'Dell 1998), while 244--440 is significantly larger,
having a similarly measured width of 3.5\arcsec.  Since the objects
all show symmetry to a greater or lesser degree about lines pointed
towards their ionizing star (\thC\ for the first three proplyds and
\thA\ for 244--440), observations were made with position angles along
and perpendicular to these lines.  Small deviations in angle were made
to avoid bright stars and an additional angle was added for 170--337
to lie along the known direction of a microjet and associated shock
feature (O'Dell 1998, Bally \etal\ 1998a).  The PA's used were 151,
231, and 353\arcdeg\ for 170--337, 135 and 225\arcdeg\ for 177--341,
62 and 152\arcdeg\ for 182--413, 52 and 142\arcdeg\ for 244--440.  The
slit PA was held fixed throughout the exposures of 300, 450, and 900
seconds by the Keck I image rotator, whose accuracy was checked using
stars of known orientation.  The astronomical seeing was typically
1.5\arcsec\ and the transparency varied from clear to partly cloudy.

Producing easily usable spectra from the complex echelle spectrograms
required a series of steps.  A 512 pixel long segment centered on the
emission line was sampled and the correct echelle order isolated.  The
spectra in these samples are tilted owing to the use of a
cross dispersion.  The tilt was removed by using modifications of
tasks within IRAF\footnote{IRAF is distributed by the National Optical
  Astronomy Observatories, which is operated by the Association of
  Universities for Research in Astronomy, Inc.\ under cooperative
  agreement with the National Science foundation.}, the result being
two dimensional spectra with the calibrated wavelength as the
\(x\)-axis and the angular position along the slit as the \(y\)-axis.
These steps were duplicated for each of the following spectral lines:
\SII, \NII, \Ha\ 6563\AA, \SIII, \OI, \HeI, \OIII, \Hb\ 4861\AA.  The
\SIIIshort\ and \OIshort\ lines were sufficiently close that they were
processed as a single sample.  Radial velocities were calculated using
the rest wavelengths employed by Esteban \etal\ (1998), with the
exceptions that a rest wavelength of 6300.31\AA\ was used for \OIshort\
according to the arguments of O'Dell \&\ Wen (1992) and 5875.74 for
\HeI, which gives the same velocity as the \OIIIshort\ lines which should
arise from the same emitting layer.  In those sections where adjacent
spectral orders overlapped there was an enhanced signal due to the
continuum of the contaminating order being added to the continuum of
the order containing the emission line of interest.  This small
contamination was subtracted from the region of overlap by appropriate
scaling.  We made certain that there was no contamination of the
subject emission lines by other emission lines in adjacent orders.

\section{Data Analysis}
\label{sec:Data-Analysis}

The primary challenge in extracting proplyd spectra from groundbased
telescope spectrograms is in the accurate subtraction of the nebular
background.  The peak surface brightness of the proplyds is higher
than that of the nebula but at typical ground observatory angular
resolutions the image is smeared and the observed surface brightness
drops to a value comparable to or less than the background, so that
one is extracting a smaller signal from a larger.  This problem is
compounded by the fact that the nebular background to be subtracted is
not homogenous and shows significant variations on scales of several
arcseconds.  At the signal levels obtained in the present observations
the uncertainty due to photon statistics was much smaller than the
uncertainties encountered in the background subtraction.  Although
most of the nebular emission occurs in a small velocity range
corresponding to emission from material accelerating away from the
main IF, there is considerable information in the line profiles near
the systemic velocities, so that accurate correction for the
background needs to be made at all velocities.  The best previous
attempt at extracting proplyd spectra is that of Henney \etal\ (1997),
who obtained profiles of five objects, with only that of 158--323 (LV5)
being good enough to make a detailed comparison with the expectations
of the models.  Even for 158--323 there was considerable uncertainty
about the profile near the systemic velocity of the nebula.  Moreover,
that investigation treated only the \OIIIshort\ 5007\AA\ line and in only
one spectrum for each object.  Therefore, it was impossible to
quantitatively assess the uncertainties of the extracted profiles.
The present data set is much more complete since it looks at emission
lines representing a broad range of ionization states and contains
multiple spectra taken at a variety of PA's.  The range of ionization
states should be useful in the diagnosis of the proplyds and the
wealth of spectra should allow determination of the errors that occur
because of vagaries in extraction of the background.

Data is not available for all lines in all proplyds.  The \Ha\ line
was saturated in all the long exposures except for 244--440.  Some
lines are intrinsically weaker and the background less well defined.
There was an object-to-object variation in the contrast of the proplyd
against the background due to the fact that the brightness of the
nebular emission, mainly arising near the IF, varies with angular
distance from the photoionizing star in a manner different from that
of the proplyds, owing to the different geometrical factors involved.
The data set used in the analysis is a subset of the total, our having
rejected saturated and low contrast emission lines.  Within this
subset some portions of the spectra were not used because of the known
presence from images of shock or jet features which would have
rendered background subtraction more uncertain.

\subsection{Subtraction of the nebular background}
\label{sec:subtract}

\placefigure{fig:example-samps}

Line profiles of the proplyds were extracted in two quite different
fashions, the two methods having their own specific advantages.  The
first method used larger background samples at slightly greater
distances from the proplyd, but employed known surface brightnesses on
\HST\ images to scale to the position of the proplyd.  The second
method relied on a simple subtraction of the background using smaller
and slightly closer samples, with the judgement about where the sample
should be taken being based on the profile along the slit.
Henceforth, these will be referenced as the ``large-sample'' and the
``small-sample'' methods.  In both methods, the continuum emission
(arising from free-free and free-bound processes in the ionized gas
plus dust scattering of starlight) was removed from each sample before
subtraction.  Examples of the samples used in the two methods are
shown in Figure~\ref{fig:example-samps}.

In the ``large-sample'' method of extraction the characteristic sample
size for the proplyd was 3.8\arcsec\ with background samples of about
2.0\arcsec\, centered at distances of about 3.0\arcsec\ from the
proplyd.  The exact sizes and distances varied according to where the
proplyd was located within the entrance slit on a particular spectrum.
With the seeing disk size that applied during the observations, these
sizes and distances allowed reasonable isolation of the smallest three
proplyds, but did sometimes subtract part of the proplyd spectra, this
problem being greatest for 244--440.  Such partial subtraction should
not alter the nature of the resultant profile, only reduce its total
signal.  Whenever possible, two background samples were taken, one to
each side of the proplyd sample, as illustrated in
Figure~\ref{fig:example-samps}.  We designate as a single profile the
result of subtracting the background based on either of the two
background samples, which means that two subtracted profiles could
result from a single spectrum.  The amount of background signal
subtracted was not simply the value in each region, rather, it was
this value scaled according to relative surface brightnesses on \HST\ 
images.  We extracted samples of the \HST\ images made in the vicinity
of each proplyd in either the same line or a line of similar
ionization (O'Dell \&\ Wong 1996) and rotated these samples to have
the same orientations as the slits employed.  Regions accurately
matching the background samples were then identified and measured.
For comparison of these background surface brightnesses to that at the
proplyd, we then identified and measured rectangular (long axis
parallel to the slit) samples close to and on both sides of the
proplyd image.  The average of those two samples was then taken to be
the surface brightness at the proplyd position.  The ratio of this
value to that of each background sample was then used to scale the
background prior to subtraction from the nebula.

The attraction of this method of background subtraction is that it
uses the best information in scaling the amount of background
subtracted and allows the isolation of the maximum amount of proplyd
spectrum.  The primary disadvantage of the method is that neither the
high, low, or proplyd background samples have been smeared according
to the astronomical seeing that applied, which varied from 1 to
2\arcsec.  The importance of this limitation is revealed in the
scatter of profiles obtained using similar sample regions on different
spectra.

In an effort to overcome some of the difficulties that arose in the
``large-sample'' method, a second attempt to extract the line profiles
was made using smaller samples (0.4--1.2$\arcsec$). The motivations
for choosing small proplyd samples were, firstly, to maximize the
ratio of proplyd to nebular emission in the sample, and, secondly, to
minimize the contribution of emission from the proplyd tail to the
sample.  The motivations for choosing small background samples were,
firstly, to ease the avoidance of ``pathological'' regions of the
nebula (for example, the bowshock that lies in front of some of
the proplyds), and, secondly, to allow the background samples to be as
close as possible to the proplyd, thereby minimizing the distance over
which the nebular profile need be interpolated.  In order to help
choose the positions of the samples, graphs were constructed of 
three parameters of the line profile (total intensity, mean velocity,
and velocity width) as a function of position along the slit.  An example
plot is shown in Figure~\ref{fig:example-samps}.  By means of these
and the \HST\ images, samples were chosen covering the peak of the
proplyd emission, together with nearby background samples that avoided
shocks and other ``features'' in the nebular background.  In this
``small-sample'' method, no attempt was made to scale the nebular
samples using the \HST\ images before subtraction from the proplyd
sample.  

\section{Extracted Proplyd Line Profiles}
\label{sec:Extr-Profiles}

\placefigure{fig:line-profiles}

The proplyd emission line profiles that result from application of the
two nebula subtraction methods of the previous section are shown in
Figure~\ref{fig:line-profiles} for the four proplyds 170--337,
177--341, 182--413, and 244--440.  Prior to the averaging of the
individual extracted profiles discussed in the previous section, these
were shifted to a common heliocentric velocity scale and normalized
using the wings of the profiles, as far as possible from the
velocities showing substantial nebular emission.  The resultant
averaged extracted profiles are of varying quality.  For the proplyds
177--341 and 244--440, the results are very good for all lines, with
the two subtraction methods agreeing to within their mutual error
bars.  For 170--341 and for the high-ionization lines of 182--416, the
results are generally poorer.  In the case of 170--341, this is mainly
because of the strongly fluctuating nature of the background nebular
profile at that position, while, in the case of 182--416, it is due to
the poor contrast between the proplyd and the nebula, especially in
high-ionization lines such as \OIII.

The ``small-sample'' method generally uses a larger number of
background samples and often has smaller scatter than the
``large-sample'' technique.  However, we caution that this does not
necessarily imply that the results of this method are more reliable.
Rather, we take the conservative view that the line profiles can only
be fully trusted where the two methods are in agreement.

\subsection{Extinction by dust in the proplyd}
\label{sec:extinct}

One unknown factor that may introduce systematic errors in the
extracted profiles is the extinction of the background nebular
emission by dust in the proplyd.  If this were a significant effect it
would result in the over-subtraction of the nebular line and could
even cause the extracted profile to become negative.  It can be seen
that for some lines the extracted profile is indeed negative for
certain velocity ranges (Fig.~\ref{fig:line-profiles}). However, it is
not clear if this is a real effect or merely due to errors in the
estimation of the nebular emission at the proplyd position.

\HST\ images of 182--413 show that the circumstellar disk in this
object is completely opaque at visible wavelengths (Bally \etal\ 
1998).  This disk is much smaller than the seeing disk in the current
observations and will reduce the nebular emission at the proplyd
position by at most 1\% (in the extreme case in which there is no
nebular emission from in front of the proplyd), which is much less
than the uncertainty due to the spatial variation of the nebular
emission.

A potentially more important effect, although one very difficult to
quantify, is extinction by dust in the extended neutral envelope of
the proplyd and in the ionized photoevaporated flow itself.  Even a
moderate optical depth here can have a significant effect if the area
of the proplyd is comparable to, or greater than, that of the seeing
disk, which is certainly the case for 182--413 and 244--440.  Henney
\& Arthur (1998) find hydrogen column densities through the ionized
photoevaporated flows in the range
\(10^{20}\)--\(10^{21}\,\mbox{cm}^{-2}\), while the models of
St\"orzer \& Hollenbach (1999) predict column densities through the
{\em neutral\/} photoevaporated flows of approximately \(5\times
10^{21}\,\mbox{cm}^{-2}\). Hence, dust in the photodissociated flow
inside of the IF will probably be the dominant factor in the
extinction of background nebular emission. 

If the dust-gas ratio of the grains responsible for the visual
extinction is ``normal'' (corresponding to an extinction cross-section
per H atom of \(5\times 10^{-22}\,\mbox{cm}^{-2}\)), then the mean
extinction optical depth through the proplyd will be about 4. However,
the ``effective'' extinction will be less than this for three possible
reasons. Firstly, apart from 244-440, all the proplyds are smaller
than the seeing width (at least across their minor axis), so the
extinction will be ``diluted'' by a factor roughly equal to the ratio
of the projected area of the proplyd to the area of the seeing disk.
In 177-341, for example, this amounts to about a factor of 3.
Secondly, some of the nebula emission may arise in front of, rather
than behind, the proplyd, and so will suffer no extinction due to dust
in the proplyd.  This is unlikely to be an issue for proplyds near
\thC, where the majority of the nebular emission is confined to a thin
layer near the principal IF (Baldwin \etal\ 1991; Wen \& O'Dell 1995),
but may become important for the more distant proplyds (182--413,
244--440). Thirdly, the nebula subtends a large solid angle as seen
from the proplyds and so, for reasonable values of the dust albedo and
scattering phase function, scattering into the line of sight will be
important (Henney 1998), possibly reducing the effective extinction by
a factor of 2 or more.  Furthermore, there exists the possibility that
the grains responsible for producing the visual extinction will be
selectively depleted in the proplyd photoevaporating flows. As
noted by Hollenbach, Yorke \& Johnstone (1999), theories of dust
settling and coagulation in disks (Weidenschilling 1984;
Weidenschilling \& Cuzzi 1993) imply that larger grains settle to the
disk midplane more rapidly than smaller grains, where they will
coagulate into larger bodies.  Hence, it may be that the \(\sim
0.1\mu{\rm m}\) grains responsible for the visual opacity are depleted
in the neutral flow from the circumstellar disk, whereas the \(\sim
0.01\mu{\rm m}\) grains responsible for the FUV opacity are not.

\placefigure{fig:dust}

Figure~\ref{fig:dust} shows examples of the effects on the subtracted
line profiles of different degrees of extinction of the background
nebula by dust in the proplyd.  The thick lines show the line profile
assuming that the proplyd is totally transparent (as in
Fig.~\ref{fig:line-profiles}) while the thin lines show the effect of
assuming an effective extinction, \(\Av\), in the range 0.1--1.0. It
can be seen that the \NIIshort\ lines in the two proplyds shown are
hardly affected by the assumed value of \(\Av\), whereas the
\OIIIshort\ lines are affected substantially.  This is partly due to
the difference in wavelength between the two lines (\OIIIshort\ is
bluer, and so suffers greater extinction for a given \(\Av\)), but
mainly due to the fact that relative brightness of the proplyds
compared to the nebula is greater in \NIIshort\ than in \OIIIshort.
The apparent minimum at the center of the \OIIIshort\ lines is seen to
disappear for \(\Av \ge 0.2\), and so its real existence is doubtful.
On the other hand, it seems that \(\Av\) cannot be much greater than
0.5, since at this level of assumed extinction the nebular line starts
to show through clearly in the extracted profile, implying that the
background has been undersubtracted.

Studies of the spatial and velocity variations in the \Ha/\Hb\ ratio
(Henney \& Watson 1999) are consistent with an effective extinction
of \(\Av =\)0.1--0.2 in the proplyds, although the exact amount is
uncertain because of possible deviations from Case B emissivity. It
remains to be seen whether or not such a low value can be reconciled with
the expected column densities in the photodissociated flows without
invoking selective depletion of the grains.

\subsection{Parameters of the extracted line profiles}
\label{sec:param}

\placefigure{fig:line-parameters}
Figure~\ref{fig:line-parameters} shows the mean velocity and widths of
all the extracted proplyd spectra (open symbols), together with the
same quantities for the adjacent background nebula samples (filled
symbols). The flux-weighted mean velocity of a line profile $I(v)$ is
given by
\[
\bar{v} = \int_{-\infty}^{\infty} v I(v)\, d v \ \Bigg/
\int_{-\infty}^{\infty} I(v)\, d v \ .
\]
For the line width, we use the quantity $2.355\, \sigma$,
where $\sigma$ is the flux-weighted RMS width of the line, given by
\[ 
\sigma = \left[ \int_{-\infty}^{\infty} (v-\bar{v})^2 I(v)\, d v \
  \Bigg/ \int_{-\infty}^{\infty} I(v)\, d v \right]^{1/2} \ . 
\] 
For a Gaussian line profile, $2.355\, \sigma$ is equal to the FWHM of
the line.  However, the line profiles are often very non-Gaussian,
both for the proplyds and for the background nebula.  In such cases,
$\sigma$ is a more robust estimate of the line width than the FWHM,
especially when the line is double-peaked.  

The line widths have been corrected for the effects of thermal and
instrumental broadening by quadrature subtraction of the widths of the
respective profiles.  The instrumental width is approximately 6 \kms\ 
and an ion kinetic temperature of $8900\,$K was assumed in all cases,
corresponding to a thermal width of 20 \kms\ for the hydrogen lines.
The lines are arranged in order of increasing ionization potential
(IP), ranging from \OI, which is expected to come from neutral gas, to
\HeI\ and \OIII, which should come from the most highly ionized gas in
the \ion{H}{2}\ region.

\subsubsection{Trends in the lines from the background nebula}

The mean velocities of the nebular lines show a clear trend of
increasing blueshift with IP, as first reported by Kaler (1967). This
trend is commonly interpreted as being due to an ionization
stratification in the nebula (more highly ionized species are found
closer to the ionizing star), coupled with an acceleration of gas away
from the ionization front (which is seen close to
face-on).\footnote{One slight inconsistency of our data with this
  picture is that \SIIshortp\ is always blueshifted by 1--2 \kms\ with
  respect to \NIIshortp.  This is the reverse of what is expected since
  the ionization potentials of S$^0$ (10.36 eV) and S$^+$ (23.33 eV)
  are much lower than those of N$^0$ (14.53 eV) and N$^+$ (29.60 eV),
  implying that the \SIIshortp\ emission should arise from partially
  ionized regions (the IF itself and the neutral photodissociation
  region), whereas the \NIIshortp\ emission should come mainly from
  the H$^+$/He$^0$ zone.  This discrepancy could be resolved if the
  rest wavelength of one or other of the two lines were in error by
  $0.02$\AA, which is well within the uncertainty of the determination
  of the rest wavelengths.} Such a scenario (Zuckerman 1973), in which
the majority of optical line emission at small angular displacements
(\(< 2\arcmin\)) comes from a thin layer close to the IF on the
surface of the background molecular cloud, has been shown to be
broadly consistent with a mass of observational material (e.g.\ 
Baldwin \etal\ 1991; O'Dell \etal\ 1993a; Wen \& O'Dell 1995).
However, a detailed physical model is still lacking.

The thermal-corrected width of the nebular emission lines does not
seem to vary significantly with IP, except for \OI, which is
consistently narrower.  However, the width does vary substantially
from position to position, being much larger at the position of
244--440.  It should be noted that the widths given here are for the
line as a whole, rather than for individual Gaussian components, as
have been used in some previous studies (Castan\~eda 1988; O'Dell \&
Wen 1992; Wen \& O'Dell 1993).

\subsubsection{Trends in the lines from the proplyds}

For three of the four proplyds, the mean velocities of the extracted
lines follow the same general trend as the nebular lines, but with the
pattern shifted 1--2 \kms\ to the blue (177--341 and 244--440) or to
the red (170--337). There is also a tendency for the velocity gradient
to be shallower for the proplyds than for the background nebula.  For
182--413, on the other hand, the trend is in the opposite direction
(lines from more highly ionized species are more redshifted).  For all
objects except 170--337 the proplyd \OI\ line is significantly
blueshifted by 3--7 \kms\ with respect to the nebula.  Possible
explanations for this are outlined in
section~\ref{sec:Kinem-proplyd-stars}.

The thermal-corrected width of the extracted proplyd lines is perhaps
the most interesting and important property plotted in
Figure~\ref{fig:line-parameters}.  In marked contrast to the behaviour
of the background nebular lines, the proplyd lines show a strong
increase in width with increasing ionization potential.  This can be
very clearly seen in three of the objects, although it is less
apparent in 182--413 because the data for the high ionization lines
in this object are very poor due to low contrast against the nebula.

This correlation can be directly understood in terms of ionization
stratification in an accelerating photoevaporating flow, in a similar
way to the velocity-IP correlation for the nebular lines.  There are
two differences, however, between the flow from the proplyds and that
from the principal IF of the nebula.  First, the proplyd flows are
much more divergent than the nebular flow, which leads to a more rapid
acceleration, especially close to the IF (Henney \& Arthur 1998).
Second, the proplyds are small compared with the scale of the
observations and their neutral portions are at least partially
transparent, which means that the observed line profiles come from gas
moving in a wide range of directions with respect to the line of
sight, both approaching and receding.  With the nebula, on the other
hand, the line profile from a given point samples only a thin pencil
beam through the emitting gas, all of which is probably moving in a
similar direction.  These are the two reasons why the proplyds show
such a spectacular linewidth-IP correlation, while the background
nebula does not.

\section{Photoevaporating Flow Models}
\label{sec:models}

\placefigure{fig:cartoon}

The photoevaporating flow models employed in this work are a
development of those presented in Henney \& Arthur (1998), building on
the previous work of Dyson (1968) and Bertoldi (1989).
Figure~\ref{fig:cartoon} shows the principal features of the model in
schematic form.  It is supposed that a strong D or D-critical
ionization front (IF) surrounds the proplyd's neutral envelope
(probably a slow photodissociated wind from the circumstellar
accretion disk, Johnstone \etal\ 1998) and that the newly ionized gas
flows freely away from the proplyd.  The ionization front on the
``front'' side of the proplyd, which faces the ionizing star, is
idealized to be hemispherical and the gas is assumed to flow radially.
The ionized gas is assumed to be isothermal, in which case the radial
profiles of density and velocity in the flow follow from Bernoulli's
equation. It is found that pressure gradients in the ionized gas cause
the flow to accelerate away from the IF, with the acceleration being
greatest in the D-critical case (when the flow leaving the IF is
exactly sonic).  The variation of the density of ionized gas with
angle around the IF is calculated assuming ionization equilibrium.
The density is highest at the point on the IF closest to the ionizing
star (sub-stellar point) and in the simplest case (no dust, no diffuse
ionizing field, and few ionizing photons reaching the IF) falls off as
$\cos^{1/2}\theta$, where $\theta$ is the angle between a point on the
IF surface and the sub-stellar point.  Note that in this approximation
the ionized density falls to zero at $\theta = 90^\circ$.

The main development of the models with respect to Henney \& Arthur
(1998) is the treatment of the diffuse ionizing field and the proplyd
tails.  This work is only summarised here and will be described in
greater detail elsewhere (Henney 1999). The diffuse ionizing
field in the Orion nebula arises mainly from the radiative
recombination of hydrogen to its ground state, with minor
contributions from helium recombination and scattering by dust
particles.  We characterize this diffuse field by the parameter
$\beta$, which denotes the ratio of direct (stellar) to diffuse
(nebular) ionizing flux at the proplyd position and we assume that the
diffuse field is isotropic.  In the case where diffuse ionizing
photons are always reabsorbed close to their point of emission
(on-the-spot approximation), one finds that $\beta\simeq 0.15$.
However, the mean-free-path of ionizing photons can be very large in
the interior of an \ion{H}{2}\ region and simple calculations suggest
that $\beta$ should lie in the range 0.01--0.05 at the positions of
typical proplyds.\footnote{Note, however, that the on-the-spot
  approximation is employed in treating the diffuse photons emitted in
  the photoevaporating flow itself.} The diffuse ionizing field
results in a non-zero value for the density at $\theta = 90^\circ$ and
also allows the ionization of the ``back'' part of the proplyd, which
the front-facing ionization front shadows from direct stellar
radiation.  This gives rise to the proplyd tails, which in the current
models are assumed to be cylindrical, with axis pointing directly away
from the ionizing star.  The ionized density is calculated as a
function of position along the tail, again assuming ionization
equilibrium, and simultaneously solving for the radiative transfer of
the direct and diffuse ionizing photons.  It is also assumed that the
ionized flow from the tail follows cylindrical radial streamlines.

Simulated images and spectra of the models are produced by calculating
the emissivity of various emission lines at each point in the ionized
flow.  Permitted lines are assumed to be due only to recombination and
forbidden lines only to collisional excitation.  Collisional
deexcitation of forbidden lines is taken into account using the
polynomial fitting formulae of Mellema (1993).  Extinction by dust in
the proplyd itself is treated, but the scattering of emission lines
into the line of sight (Henney 1998) is not included.  The ionization
structure of the photoevaporating flow is not calculated
self-consistently.  Instead, the position and thickness of the He
ionization front are treated as free parameters of the model.

All models presented in the following sections were calculated using
an ionized gas temperature of \(8900\,\mbox{K}\) (Liu \etal\ 1995).
It seems likely that the temperature will be slightly higher for
positions close to the IF, both because of the hardening of the
ionizing radiation field and because of the reduced cooling efficiency
at higher densities.  However, models calculated using such a
temperature profile lead to qualitatively similar results to the
isothermal models. Such models are not discussed further here because
of the increase in the number of model parameters that they entail.

\section{Model fits to \HST\ imagery}
\label{sec:HST-imagery}
\placefigure{fig:WFC} 

In order to restrict the range of models that we try to fit to the
emission line spectra, we have used \HST\ images of the proplyds to
fix as many model parameters as possible.  As an example,
Figure~\ref{fig:WFC} compares Wide Field Camera (WFC) images of
177--341 with simulated model images in three different emission
lines.  \Ha\ images of the heads of the proplyds allow one to
determine with reasonable certainty both the radius, $r_0$, of the
forward-facing IF and the peak density, $n_0$, in the ionized
photoevaporated flow. (Estimation of this density requires that
the observed surface brightness be corrected for foreground
extinction.) These parameters could be estimated for all four objects
and are listed in Table~\ref{tab:HST}, together with the foreground
extinction to each proplyd.  One can also estimate the inclination
angle, $i$, of the proplyd axis to the line of sight (Henney \& Arthur
1998), but this is rather uncertain and one cannot distinguish between
a proplyd pointing towards or away from the observer. Additionally,
estimation of $i$ was impossible in 244--440 (because of the probable
influence of two ionizing stars) and in 170--337 (because of the
microjet). 

The relative brightness in \Ha\ of the proplyd tail with respect to
the head allows one to determine the ratio of diffuse to direct
ionizing flux, $\beta$, although this is somewhat dependent on the
assumed value of $i$.  Similarly, one can measure the quantity $l_0
\sin i$, where $l_0$ is the length of the proplyd tail.  Note that the
model line profiles shown in the next section are insensitive to the
assumed values of $\beta$ and $\l_0$, especially the latter.  This is
because the emission from the brightness peak of the proplyd, where
the contribution from the tail is small, has been isolated in the
extracted proplyd line profiles.  By fitting to the \NIIshort\ and \OIIIshort\ 
images, one can then estimate the radius, $r_1$ and thickness, $\Delta
r_1$, of the He ionization front (assumed to coincide with the ${\rm
  N}^+/{\rm N}^{++}$ and ${\rm O}^+/{\rm O}^{++}$ boundaries).

\newcommand{\Sp}{\rule[-1.5ex]{0pt}{4ex}}
\newcommand{\PM}[2]{{\scriptscriptstyle +#1 \atop
    \raisebox{.5ex}{\(\scriptscriptstyle -#2\)} }}
\begin{table*}[tbp] \IfPPelse{\small}{\scriptsize}
  \caption{Proplyd parameters derived from \HST\ images and Keck spectra
    \label{tab:HST}} 
  \begin{tabular}[t]{cccccccc ccc}\hline
    \Sp & & \multicolumn{6}{c}{\dotfill~\HST~\dotfill} &
    \multicolumn{2}{c}{\dotfill~Keck~\dotfill} & \\
     &  & 
    \(r_0\tablenotemark{b}\) &
    \(n_0\tablenotemark{c}\) & \(i\,\tablenotemark{d}\) & 
    &  & & \(i\,\tablenotemark{d,h}\)& \(u_0\tablenotemark{g,h}\) & \(\dot{M}\tablenotemark{i}\)
     \\ 
    \Sp Proplyd  & \(\tau_{\rm f}(\mbox{H}\alpha)\,\tablenotemark{a}\) & 
    (\(10^{15} \mbox{cm}\)) & (\(10^6\mbox{cm}^{-3}\)) &
     (degrees) &\(\beta\,\tablenotemark{e}\) &\(r_1/r_0\tablenotemark{f}\)&\(\Delta r_1/r_0\tablenotemark{f}\) &
    (degrees) & (\(\kms\)) & (\(10^{-7}\smy\)) \\
    \hline
    \Sp 170--337 & 1.43 & \( 1.17 \) & \(1.31\) &
    --- & --- & --- & --- & \( < 90\) & \(\sim 13\) & \( 8\PM{5}{3} \) \\
    \Sp 177--341 & 1.38 & \( 1.91 \) & \(0.636\) & 
    45--135 & 0.01 & 1.1 & 0.1 & 75--85 & 13--16 & \(9 \pm 4\)\\
    \Sp 182--413 & 1.09 & \( 3.69 \) & \(0.153\) &
    0--30, 150--180& --- & --- & --- & \(> 90\) & \(\sim 13\) & \(7
    \pm 5 \)\\
    \Sp 244--440 & 0.51 & \( 10.4 \) & \(0.0785\)  &
    --- & --- & 1.25 & 0.2 & \( <90\) & \(\sim 13\) & \(15 \pm 7\) \\  
    \hline
  \end{tabular} \\[1ex]
  {\scriptsize \parbox{\textwidth}{
      \tablenotetext{a}{Foreground dust extinction optical depth at
        \Ha, derived from measurements of the Balmer decrement at
        adjacent nebular positions (O'Dell, Walter, \& Dufour
        1992; O'Dell 1998).} 
      \tablenotetext{b}{Radius of forward-facing IF assuming a distance of
      430 parsecs.}
      \tablenotetext{c}{Peak density in photoevaporating flow.}
      \tablenotetext{d}{Ranges of possible inclination angles of proplyd
      axis. Note that separate columns are given for the constraints
      from imaging and from spectroscopy.}
      \tablenotetext{e}{Ratio of diffuse to direct ionizing flux.}
      \tablenotetext{f}{Relative radius and thickness of He IF.}
      \tablenotetext{g}{Initial velocity of newly-ionized gas leaving IF.}
      \tablenotetext{h}{For 177--341, these parameters are determined
        from the detailed model fits to the emission line profiles
        (Section~\ref{sec:Model-fits}). For the other objects, they
        are estimated from the behaviour of the mean velocity and
        velocity width (Fig.~\ref{fig:line-parameters}).}
      \tablenotetext{i}{Derived mass loss rate.}
      }}
\end{table*}

For 177--341 it can be seen (Figure~\ref{fig:WFC}) that the models are
very successful in reproducing the \HST\ images in the three lines,
especially the head of the proplyd.  Although the reduced $\chi^2$ of
the fits is typically in the range 5--20, this is mainly because of
the small but significant deviations from cylindrical symmetry of the
proplyd, which obviously cannot be captured in the current models.
The quality of the fit near the tips of the cusp is much improved with
respect to the models presented in Henney \& Arthur (1998), which did
not include the effect of the diffuse ionizing radiation.  Although
the fits to the tail well reproduce the intensity profile along the
proplyd axis (top panels of Figure~\ref{fig:WFC}), the model images
have tails that are significantly wider than observed, indicating that
the tails are not the cylinders assumed in the model, but instead
taper towards their tip.  The ionization stratification in the proplyd
is readily apparent from the increase in size of the images as one
passes from \NIIshort\ through \Ha\ to \OIIIshort.  The best-fit models have
$r_1/r_0 = 1.1$, so that radius of the ${\rm N}^+/{\rm
  N}^{++}$ and ${\rm O}^+/{\rm O}^{++}$ transition (presumably
corresponding to the He ionization front) is about 10\% larger than the
radius of the H ionization front.  One also needs quite a broad ${\rm
  N}^+/{\rm N}^{++}$ transition ($\Delta r_1/r_0 = 0.1$) in order to
reproduce the intensity profile of the \NIIshort\ image. 

For the other three proplyds, although the model fits to the heads of
the proplyds are satisfactory in all cases, the fits to the tails are
much less convincing than in the case of 177--341.  This is mainly
because the assumption of a cylindrical tail is an even poorer
approximation for these objects, which all show tails that are more
conical in form.  Additionally, 182--413 and 244--440 both show
substantial deviations from cylindrical symmetry with respect to the
direction of the ionizing star.  In the case of 244--440, this is
probably because the proplyd is influenced both by \thC\ and \thA.
The same may be true of 182--413, although in this object it is only
the tail which shows the asymmetry.  Alternatively, the tail in this
object may be influenced by the orientation of the circumstellar disk,
either via an anisotropic flow from the disk or via magnetic field
lines (Bertoldi \& Johnstone 1998).  As a result, the parameter
$\beta$ can only be reliably determined for 177--341.  However, the
deficiencies of the photoevaporating flow models in explaining the
tails of these objects should not greatly affect the predicted line
profiles for the reasons advanced earlier in this section.  The
position and thickness of the He ionization front could only be
determined for 177--341 and 244-440 since in 182--413 the \OIIIshort\ shell
is too irregular to reliably determine its radius of curvature, while
in 170--337 the presence of a microjet complicates the interpretation
of the images.

\subsection{Validation of densities derived from model fits}
\label{sec:Valid-dens-deriv}

To obtain the peak densities, \(n_0\), listed in Table~\ref{tab:HST},
the extinction-corrected \Ha\ intensities are converted into
emission measures assuming that the \Ha\ emission coefficient is
known. The ionized density at the IF on the proplyd axis is then found 
using the density profile predicted by the photoevaporating flow
model. Although this approach seems to be robust, it is worthwhile to
try and validate the densities so obtained by an independent method. 

Unfortunately, the ionized densities in the proplyd flows are so large
(\(10^5\)--\(10^6\,\mbox{cm}^{-3}\)) that the \SIIshort\ 
6716\AA/6731\AA\ doublet ratio, which is traditionally used as a
density diagnostic in ionized regions, will be in the high-density
limit. Hence, it is necessary to find a diagnostic pair of lines with
a higher critical density. Just such a pair exist in the ion
C\(^{++}\): the magnetic quadrupole transition [\ion{C}{3}]\ 1907\AA\ 
and the ``semi-forbidden'' electric dipole transition \ion{C}{3}]\ 
1909\AA\ (Osterbrock 1989, p.\ 136). Although these lines were
observed in two proplyds with the Faint Object Spectrograph (FOS) of
the \HST\ (Bally \etal\ 1998a), the FOS spectral resolution
(\(R=1300\)) is not sufficient to fully resolve the doublet. We are
therefore grateful to Robert Rubin for allowing us access to
unpublished FUV longslit spectra of the Orion nebula obtained with the
Space Telescope Imaging Spectrograph (STIS), which include emission
from the proplyd 159--350 (HST 3).
 
After subtracting the nebular contribution to the emission, we
determine a value of 0.4 for the ratio [\ion{C}{3}]\ 
1907\AA/\ion{C}{3}]\ 1909\AA, which corresponds to an electron
density of \(1.01\pm 0.05 \times 10^5\,\mbox{cm}^{-3}\) for
temperatures in the range \(8000\)--\(10000\,\mbox{K}\). Henney \&
Arthur (1998) determined a peak density of \(n_0 = 6.5 \times
10^5\,\mbox{cm}^{-3}\) for the same proplyd. However, in order to
meaningfully compare these values, two factors must be taken into
account: first, that  C\(^{++}\) will only be present in the
regions of the flow where helium is ionized, and, second, that the
STIS observations sample an extended region of the proplyd containing
a range of ionized densities. Both these effects should tend to make
the observed C\(^{++}\) density less than \(n_0\).

In order to quantitatively assess these effects, we employed model
fits to the \HST\ PC images of 159-350 in \Ha\ and \OIII\ in order to
determine the radius of the C\(^{++}\)/C\(^{+}\) transition in terms
of the IF radius, \(r_0\) (assuming that the ionization of carbon
follows that of oxygen).  Unfortunately, the analysis is complicated
by the fact that 159--350 is a binary system (projected separation
0.5\arcsec) in which both components are proplyds. The smaller and
fainter of the two proplyds, which lies between the larger proplyd and
\thC, seems to shadow part of the IF of the larger and brighter one,
leading to a marked asymmetry in the appearance of the latter,
especially in \OIIIshort.  As a result, we can only constrain the
radius of the C\(^{++}\)/C\(^{+}\) transition to lie in the range
1.2--1.4\(r_0\).  Using these values, we than calculate from our
models the emissivity-weighted mean density in the C\(^{++}\) zone in
an aperture corresponding to the STIS observations.  For a peak
density of \(n_0 = 6.5 \times 10^5\,\mbox{cm}^{-3}\), this density
turns out to be \(0.9\)--\(1.1 \times 10^5\,\mbox{cm}^{-3}\), in
excellent agreement with the observed value given above. 

The concordance we find in this object between the densities derived
from our model fits and those obtained from the C\(^{++}\) doublet
give us confidence that the model-derived densities are indeed
accurate.

\section{Model fits to the proplyd line profiles}
\label{sec:Model-fits}

Although the methods discussed in the previous section allow one to
fix many of the model parameters by comparison with \HST\ images of
the proplyds, there are some parameters which are not very well
constrained in this way.  Of these, the inclination, \(i\), and the
initial Mach number, \(M_0\), of the photoevaporating flow are the
parameters that have most effect on the predicted model line profiles.
Hence, by comparison of model predictions with the observed line
profiles one can both check the validity of the photoevaporating flow
model and try to determine these two parameters.  In this section, we
concentrate on the proplyd 177--341, both because the data quality of
the observed line profiles are highest in this object and because the
\HST\ images show a high degree of consistency with the simple
photoevaporating flow models employed.  The data quality is also high
for 244--440, but in this case the interpretation is complicated by
the probable influence of two ionizing stars.  This object, which is
also the only object to be clearly resolved spatially at ground-based
resolutions, will be discussed further in a subsequent paper.  For the
remaining two proplyds, 170--337 and 182--413, the quality of the
extracted spectra is too low to meaningfully derive parameters from
the model fits.

\placefigure{fig:fit177}

Figure~\ref{fig:fit177} shows the results of fitting
such models to the observed line profiles.  The fitting procedure is
to perform a \(\chi^2\) minimization of the model spectrum for various
values of $i$ and $M_0$, in order to determine the best-fit values of
the profile normalization and velocity zero point (corresponding to
the heliocentric velocity of the proplyd central star).  At each step,
the model is convolved with the instrumental profile and seeing before
extracting a spectrum from an aperture identical to that used in the
observations.  Results are shown in the figure for models with
$M_0=1$.  For each emission line, the figure shows the fit for the
value of $i$ that gives the lowest value of \(\chi^2\), plus the
highest and lowest values of $i$ that give an acceptable fit (taken as
\(\chi^2/\nu < 3\), where \(\nu\) is the number of degrees of
freedom).  Also shown in the figure is the best-fit static model, in
which the line broadening is due solely to the thermal Doppler effect
(together with fine-structure splitting in the case of \HeI).

It can be seen that the models fit quite well and that the inclination
of this object is well constrained to lie in the range \( i = 75
\)--\( 85^\circ \), consistently between the different emission lines.
The derived velocity zero point (\(\simeq 22\)\kms) is also similar
for all the lines and is consistent with the centroid velocity of the
\OI\ emission.  The static model completely fails to fit the observed
line profiles.  Similarly acceptable fits are found for all $M_0 <
1.2$.  

The mass loss rate of the photoevaporating flow in the best-fit model
is \( 9.4 \times 10^{-7} \mbox{M}_\odot \, \mbox{yr}^{-1} \), with the
contributions of the flows from the head and the tail being roughly
equal. This derived mass loss rate is directly proportional to \(u_0
n_0 r_0^2\), where \(u_0 = 13 M_0 \)\kms\ is the initial velocity of
the photoevaporating flow, \(n_0\) is its peak density, and \(r_0\) is
the radius of the ionization front (Table~\ref{tab:HST}), each of
which are known to an accuracy of 10--20\% (Henney \& Arthur 1998).
Additionally, the parameters of the proplyd tail, which provides half
the mass loss, are less well constrained by the fits than those of the
head. As a result, we estimate that the uncertainty in the derived
mass loss rate is of order 50\%.
 
\section{Discussion}
\label{sec:Discussion}

In this section, we critically discuss the mass loss rates determined
by us and by previous authors, together with the observational
measurements of the disk masses and the implications of these for the
length of time that the proplyds have been exposed to ionizing
radiation.

\subsection{Derived mass loss rates}
\label{sec:Derived-mass-loss}

We have performed high spectral resolution spectroscopy of four
proplyds in several different emission lines.  We have shown that the
proplyd line width increases with ionization potential, which is
consistent with the idea that the ionized gas in the proplyds
comprises an accelerating ionization-stratified flow. Detailed model
fitting to 177--341, which is the proplyd with the highest quality
data, shows that the ionization front in this object is approximately
D-critical and gives a mass loss rate in the evaporated flow of \( 9
\pm 4 \times 10^{-7} \mbox{M}_\odot \, \mbox{yr}^{-1} \).  Static
models were found to be completely unable to fit the observations

The other proplyds all show similar linewidths to 177--341
(Figure~\ref{fig:line-parameters}), implying that the flow speeds in
these objects are also similar, even in the absence of detailed model
fits to the line profiles. That being the case, one can use the
parameters deduced from model fits to the HST images
(Section~\ref{sec:HST-imagery}) to calculate the mass loss rates of
these objects by scaling the value for 177--341.  The resultant rates
are \( \dot{M} = 8 \times 10^{-7} \), \( 7 \times 10^{-7} \), and \(
1.5 \times 10^{-6} \mbox{M}_\odot \, \mbox{yr}^{-1} \) for 170--337,
182--413, 244--440, respectively.\footnote{For 244--440, the value
  obtained by scaling the result for 177--341 has been divided by two
  because of the lack of a prominent tail in this object.} Assuming
that the flow speeds are the same in all proplyds, not just those
observed in the present study, then one can calculate \( \dot{M} \)
for all proplyds that have well-determined values of \( r_0 \) and \(
n_0 \). Taking the sample from Henney \& Arthur (1998), which
comprises 29 proplyds within 30 arcsec of \thC, one then finds the
distribution of mass loss rates shown in Figure~\ref{fig:mass-loss}.
It can be seen that the four proplyds in the current sample have
higher than average mass loss rates and indeed the mean value for the
combined samples is only \( \dot{M} = 4.1 \times 10^{-7}
\mbox{M}_\odot \, \mbox{yr}^{-1} \).

\placefigure{fig:mass-loss}

\subsubsection{Comparison with previous mass loss determinations}
\label{sec:Comp-with-prev}
The mass loss rates that we derive can be compared with previous
estimates, starting with those of Churchwell \etal\ (1987). These
authors used a method similar to ours, except that they used
interferometric images of the 2 cm radio free-free emission in order
to estimate the emission measure of the ionized gas.  Furthermore,
they had to {\em assume\/} that the gas was expanding at the sound
speed (an assumption that is confirmed by the present work) and they
employed a simplified, spherically symmetric model of the ionized
photoevaporating flow. They found mass loss rates between \( 2 \times
10 ^{-7} \) and \(10^{-6} \smy \) for 6 of the proplyds closest to
\thC.  These estimates agree closely with our own derived mass loss
rates from the same objects, which is not surprising given the
similarity in methodologies.

An independent way of estimating the proplyd mass loss rates was first
employed by Johnstone \etal\ (1998) and subsequently refined by
St\"orzer \& Hollenbach (1999). This method supposes that the mass
loss is controlled by the FUV photons incident on the circumstellar
disk, which heat the disk surface and drive a photodissociated neutral
photoevaporating flow. This flow will autoregulate so that the FUV
flux that penetrates to the base of the flow is just sufficient to
heat the gas at the disk surface to a temperature such that its sound
speed exceeds the local escape velocity. The mass loss rate is then
directly proportional to the disk radius multiplied by the velocity
multiplied by the column density of the neutral flow.  St\"orzer \&
Hollenbach (1999) calculate this column density using both equilibrium
and non-equilibrium PDR codes and find only a shallow dependence on
the distance of the proplyd from \thC. They then use this result to
determine the mass loss rates for a sample of 10 proplyds for which
the disk radius has been estimated or measured.  Two of the 4 objects
in our spectroscopic sample (182--413 and 171--340) are also modelled
by them and for these objects they determine mass loss rates that are
slightly less than half the rate determined by us.  However, St\"orzer
\& Hollenbach only consider the mass loss from the directly
illuminated face of the disk, whereas the diffuse nebular FUV
radiation should also cause mass loss from the back face of the disk.
In fact, as pointed out by Johnstone \etal\ (1998), since the
mass loss rate is only a weak function of the FUV flux, the mass loss
from the back face should be nearly as large as that from the
front.\footnote{This is compatible with our finding that at least half
  the proplyd mass loss occurs through the tails in some objects,
  although it should be pointed out that material leaving the directly
  illuminated face of the disk may end up in the tail, either because
  the disk axis is not aligned with the proplyd axis, or through
  lateral pressure gradients in the shocked neutral layer between the
  disk and the IF.} Hence, taking into account the flow from both
faces would bring St\"orzer \& Hollenbach's mass loss rates into
reasonable agreement with our own, especially considering that the
velocity of the neutral flow is not very well constrained in their
models.

O'Dell (1998) suggested that a confining force acted on the proplyd
``atmospheres'', resulting in very low mass loss rates. However, our
spectroscopic results conclusively rule out this possibility.
Although only one of our proplyds was suitable for fitting to detailed
models, all of them showed excess line broadening. If this broadening
is due to free expansion, as seems likely, then this means that all of
the objects studied are undergoing significant mass loss.  The
exponential decay in the density of the proplyd atmospheres found by
O'Dell (1998) can equally well be explained by the freely expanding
models we have calculated here and the large line widths are a direct
indication that the static model does not apply.  This means that it
is not necessary to seek the operation of a confining force (O'Dell
1998), which is just as well, for we now understand that the only
force sufficiently strong to confine the atmospheres (\lya\ radiation
pressure) was incorrectly derived, with the magnitude estimated by
O'Dell (1998) being, at best, an upper limit (Henney \& Arthur 1998).
Therefore, one must look to other novel methods for explaining the
existence of circumstellar envelopes in such an intense photoionizing
radiation field.

The one caveat to our conclusion arises from our assumption that the
mechanism producing the non-thermal, non-Kolmogorov line broadening in
the main body of the nebula is not operating in the ionized atmosphere
of the proplyds.  This line broadening is about 10 \kms\ (O'Dell 1994,
1999) and is common to all lines and ions.  One explanation of the
extra broadening is that of Ferland (1999) who proposes that it is due
to Alfv\'en waves, a mechanism thought to explain the non-thermal line
broadening in molecular clouds. The ad hoc addition of such an
additional broadening of the lines would possibly reduce the mass loss
rates necessary to explain the observed broad profiles of the proplyd
spectra; but, without understanding the process in the nebula, we
cannot introduce it into our proplyd models.  If the Alfv\'en wave
broadening in 177-341 were the same as that proposed in the main body
of the nebula, then the derived mass loss rate would hardly be
reduced.  Invoking larger values for any Alfv\'en wave component seems
without justification without understanding the physics of this
process.  Besides, an extra broadening component of even 10\kms\ would
pose problems for the model fits to the \NII\ line, whose width is
currently well reproduced by the broadening due to the expansion of
the ionized gas alone.  It is possible that the extra broadening seen
in the nebula is due to the kinematics of the gas, rather than to
Alfv\'en waves (e.g.\ Yorke, Tenorio-Tagle \& Bodenheimer 1984),
although it must be conceded that such a ``kinematic'' broadening
mechanism seems incapable of explaining the observed width of the
\OIshort\ line.

\subsection{Implications for proplyd evaporation times}
\label{sec:Impl-proplyd-lifet}
Now that we have reliable estimates for the proplyd mass loss rates,
we are in a position to calculate their evaporation times if we can
only estimate the mass of the circumstellar disks contained within
them.  Discounting the mass of the star itself, the disks should
dominate the mass of the proplyd.  The calculated masses of the
combined neutral and ionized envelopes are only of order \(10^{-5}
M_\odot\), which would imply evaporation times, \(\tevap \equiv
M/\dot{M}\), as short as 10--100 years unless a substantial reservoir
of gas in the form of an accretion disk were present.  We now know
from both observations and theory that most of the mass resides in an
inner disk, which means that the depletion of that mass will determine
the survival of the proplyds.  In this section we summarise the
arguments leading to estimates of the disk masses and discuss the
problem of their predicted short lifetimes, which conflict with
their ubiquity.

Johnstone \etal\ (1998) show that for a disk whose surface density
follows a power law in radius, then \(\tevap\) calculated using the
current values of the disk mass and mass loss rate should be
proportional to the length of time since the disk was first exposed to
FUV/EUV radiation, the constant of proportionality being unity if the
mass loss is controlled by the FUV radiation and if the surface
density follows the standard \(r^{-3/2}\) law (Adams, Shu \& Lada
1988).  According to this picture the disks started with much larger
sizes, masses and mass loss rates than they have today and have shrunk
to their current size due to photoevaporation, with the disk mass
declining in time as \(t^{-1}\) and the disk radius and mass loss rate
declining as \(t^{-2}\).  Projecting a proplyd's evolution into the
future, \(\tevap\) is hence the time it will take for the disk to lose
half its current mass and shrink in radius by a factor of 4.  If the
disk radius was initially truncated by some other process (Hollenbach
\etal\ 2000) before it was exposed to UV radiation, then \(\tevap\)
would only be an upper limit to the exposure time.

\subsubsection{Disk masses}
\label{sec:Disk-masses}

The masses of the circumstellar disks associated with the Orion
proplyds have been estimated by two different methods. The first
involves measuring the extinction by the silhouette disks of the
background nebular emission and gives only lower limits on the disk
mass (O'Dell \& Wen 1994; McCaughrean \& O'Dell 1996; McCaughrean
\etal\ 1998; Throop \etal\ 1998). Assuming ``standard'' dust
properties, the values found are typically \(10^{-6}\)--\(10^{-4}
M_\odot\). Throop \etal\ (1998), on the other hand, suggest that
substantial dust coagulation has ocurred in the disks, resulting in
grains larger than 10\(\mu\)m, which would imply that these lower
limits should be revised upward by several orders of magnitude.
However, this result is based on a claimed wavelength independence of
the size of the silhouette disks, which seems inconsistent with the
finding of McCaughrean \etal\ (1998) that the silhouette of the giant
disk 114--426 appears 20\% smaller in Pa\(\alpha\) (1.87\(\mu\)m) than
in \Ha\ (0.66\(\mu\)m).

The second method is to search for mm-wavelength thermal emission from
dust in the disks. The first attempt to do this was by Mundy \etal\ 
(1995), who observed a 45\arcsec\ radius region of the central
Trapezium Cluster at 3.5 mm with an angular resolution of
approximately 1\arcsec.  Although they detected emission from several
proplyds (notably, 167--317, 158--323, 168--326 and 159--350), in no
case was the 3.5 mm flux larger than the 2 cm flux measured by Felli
\etal\ (1993). This implies that the 3.5 mm emission is dominated by
free-free radiation in the ionized gas or non-thermal processes,
rather than by the thermal emission of dust in the accretion disk.
Hence, only upper limits on the disk masses could be determined.
These were \(< 0.15 M_\odot\) for individual sources and \(< 0.03
M_\odot\) for the average mass of all disks in their surveyed region.
Subsequently, Lada \etal\ (1996), using higher sensitivity
observations at 1.3 mm, reported the detection of thermal dust emision
from 3 proplyds (roughly one quarter of those present in their fields)
with deduced disk masses lying between \(0.007\) and \(0.016
M_\odot\).  More recently Bally \etal (1998b) has detected thermal
emission, also at 1.3 mm, from the giant silhouette disk 114--426,
which implies a mass for this object of \(0.02 M_\odot\), together
with an upper limit for the disk mass of 182--413 (one of our
spectroscopic sample) of \(< 0.015 M_\odot\).  These authors also
derive independent upper limits to the disk masses in these two
systems from the lack of observed \({}^{13}\)CO emission, but these
latter estimates depend critically on the CO abundance in the
disks, which is very uncertain (Dutrey \etal\ 1996).

In summary, if the results of the mm continuum studies are taken at face
value, then the disk masses range from \(0.02 M_\odot\) in the largest
silhouette disk (disk radius \(\simeq 1\arcsec\)) through \(\simeq
0.01 M_\odot\) for the largest of the bright proplyds (disk radii
\(\simeq 0.1\arcsec\)) down to upper limits of \(< 0.005 M_\odot\) for
the majority of proplyds (disk radii \( < 0.05\arcsec \)--\(
0.1\arcsec \)).  These can be compared with the masses measured by the
same technique for disks around classical T Tauri stars of similar
ages to those in the Trapezium cluster, but located in isolated star
forming regions such as Taurus, Ophiuchus and Lupus, which are in the
range \(10^{-4}\)--\(10^{-1} M_\odot\) (Osterloh \& Beckwith 1995; Dutrey
\etal\ 1996; N\"urnberger \etal\ 1997a,b).  Taking that subsample of
the disks observed by Osterloh \& Beckwith (1995) whose central stars
have estimated ages between \(10^5\) and \(10^6\) years (\(N=18\)),
one finds a mean mass of \(0.017 M_\odot\), although Hartmann \etal\ 
(1998) suggest that the Osterloh \& Beckwith masses may be
underestimated by a factor of 2.5.

Hence, there is no obvious conflict between the low masses found for
the disks around the stars in the dense Trapezium Cluster and those
found in more quiescent regions of low-mass star formation: the
largest disks in Orion have a mass typical of those around T Tauri
stars, whereas the disks inside the bright proplyds have lower masses
because of the photoevaporation-induced mass loss.  However, the
masses derived for the disks in Orion are extremely uncertain because
of their sensitivity to the disk temperature and the mm-wavelength
opacity of the dust. Many isolated T Tauri disks have well-measured
spectral energy distributions from infrared to mm wavelengths,
allowing reasonable confidence in the fitting of multi-parameter disk
models.  For the Orion disks, on the other hand, the disk masses are
estimated from measurements at a single wavelength and are hence much
less reliable.  More importantly, the magnitude and wavelength
dependence of the dust opacity is very poorly known, as is the
dust-gas ratio in the disks (Beckwith 1999).  In addition, the mass
estimates assume that the disks are optically thin, whereas at 1.3 mm
the disks should become optically thick for radii within
\(\simeq 10\mbox{AU}\) of the central star (Hartmann \etal\ 1998),
which is comparable to the outer radii of the smaller proplyd
disks.\footnote{For edge-on disks, such as in 182--413, optical depth
  effects will become important at even larger radii.}

If the disk masses derived from mm continuum observations are correct,
then the evaporation times would be \(\tevap \simeq 10^4\) years for
the larger proplyds (\(M\simeq 0.01 M_\odot\), \(\dot{M} \simeq
10^{-6} \smy\)) and \(\tevap \le 2.5\times 10^4\) years for the
smaller ones (\(M \le 0.005 M_\odot\), \(\dot{M} \simeq 2 \times
10^{-7} \smy\)).  However, as a result of the uncertainties discussed
in the previous paragraph, the quoted disk masses and hence the
estimated evaporation times can probably not be trusted to better than
an order of magnitude.

 
\subsubsection{The age of the Orion nebula}
\label{sec:Ioniz-age-nebula}

The ages of the low- to intermediate-mass stars in the Orion Nebula
Cluster range from roughly \(10^4\mbox{--}10^7\) years, as determined
by fitting evolutionary tracks to the observed color-luminosity
diagram (Hillenbrand 1997).  Taking that subsample (\(N=58\)) of the
stars in Hillenbrand (1997) that both have measured ages and are
listed as proplyds in O'Dell \& Wong (1996), one finds that the mean
and standard deviation of the logarithm of the stellar ages is
\(\log(t) = 5.45 \pm 0.95\).\footnote{It should be noted that these
  ages are somewhat sensitive to which stellar evolution calculations
  are used. The quoted values employ the calculations of D'Antona \&
  Mazitelli (1994), whereas the calculations of Swenson, \etal\ (1994)
  give ages that are typically 2--3 times larger.}  The ages of the
high-mass stars, such as \thC, are much harder to determine
observationally but, in the absence of evidence to the contrary, are
likely to lie in the same range.

The derived photoevaporation times are thus small compared with the
stellar ages, even taking into account possible errors in the disk
mass estimates.  Hence, one must look for other mechanisms that may
have saved the proplyd disks from exposure to UV radiation until
relatively recently.  One possibility, independently suggested by
Bally \etal\ (1998b) and Henney \& Arthur (1998), is that the ionized
zone around \thC\ was maintained in an ultracompact stage (e.g.\ 
Kurtz, \etal\ 2000 and references therein) for much of the lifetime of
the ionizing star and has only recently undergone an ``inflationary
phase'', in which it expanded to its current size.  The problem with
this argument is that the Orion nebula has a diameter of approximately
3 parsecs (\(10^{19} \mbox{cm}\)), as revealed by long-exposure
optical photographs.  Hence, if it were to have been much smaller
\(10^4\) years ago, its expansion velocity must be of order
\(150\kms\).  This is more than ten times the sound speed in the
ionized gas and seems implausibly high, even for an \ion{H}{2}\ region
in the champagne phase (e.g.\ Yorke 1986).  Numerical hydrodynamic
simulations (Yorke \etal\ 1984; Garc\'\i a-Segura \& Franco 1996) show
that the ionized gas can only reach velocities of order \(40 \kms\),
or \(80 \kms\) if one includes the effect of the stellar wind from the
ionizing star (Comer\'on 1997).  On the other hand, the expansion of
the IF need not necessarily involve physical movement of the gas: if
the motion of \thC\ had recently caused it to emerge from the
background molecular cloud, then an R-type IF may have rapidly
propagated out to the observed size of the nebula.  However, the
shell-like appearance of the outer boundary of the nebula would argue
against this idea.  Besides, in order for \thC\ to have travelled the
\(\simeq 3 \mbox{--} 6 \times 10^{17} \mbox{cm}\) from the background
cloud to its present position (Baldwin \etal 1991; Wen \& O'Dell 1995)
in the last \(10^4\) years, its velocity must be of order
\(10\mbox{--}20 \kms\), which is much higher than the velocity
dispersion of the stars in the cluster.\footnote{Note, however, that
  just such a high blue shift of \thC\ is implied by the study of
  Stahl \etal\ (1996), but the history of spectral variations and
  large range of derived radial velocities (O'Dell 1999) make it most
  likely that one is seeing peculiar atmospheric effects rather than
  the systemic velocity of the star. } The fact that the principal
emitting layer in the core of the nebula has an effective thickness
less than the distance of \thC\ from the IF (Wen \& O'Dell 1995) is
also more consistent with a highly evolved champagne region than one
in which the ``blowout'' occurred recently (G. Garc\'{\i}a-Segura,
priv.\ comm.).  It is possible, however, that other, less massive
stars in the region are partly responsible for ionizing the nebula on
large scales (e.g.\ \(\iota\,\)Ori, O'Dell \etal\ 1993a) and these may
be at a sufficient distance from the proplyds that they have induced
little disk evaporation themselves.

A further argument against the recent expansion of the \ion{H}{2}\ 
region is the statistical unlikelihood of catching \thC\ during such a
short-lived phase of its evolution.  This statistical argument also
weighs against the related proposal (Bally \etal\ 1998b) that \thC\ 
has only recently reached the main sequence and begun emitting
ionizing photons.  A similar argument would seem to rule out the
suggestion (S. Lizano, priv.\ comm.) that massive circumstellar
envelopes have protected the disks from evaporation until recently,
given that none of the bright proplyds show any surviving trace of
such an envelope.\footnote{Although one of the bright proplyds,
  244--440, is very much larger than the others, the visibility of the
  central star implies that any extended envelope in this object
  cannot be more massive than about \(10^{-4} M_\odot\).}

\subsubsection{Kinematics of the proplyd stars}
\label{sec:Kinem-proplyd-stars}

Since the stars move inside the potential well of the cluster, it may
be that the stellar motions themselves are the limiting factor in
determining the exposure time to evaporation of the proplyd disks.
Jones \& Walker (1988) calculated the proper motions of hundreds of
ONC stars and found a one-dimensional velocity dispersion of \(\sim
2.5 \kms\), which hardly varies with radius in the cluster.  Such an
``isothermal'' behaviour of the velocity dispersion is consistent with
the derived stellar density distribution (Henney \& Arthur 1998),
which is close to \(r^{-2}\) outside of a core radius of \(10\mbox{--}
15\arcsec\) around \thC , which lies very close to the center of the
cluster.  Additionally, Jones \& Walker show that the stellar
velocities are closely isotropic in all but the outermost regions of
the ONC, which means that the above velocity dispersion should also be
a typical velocity along the radius joining the star to the cluster
center.

The effect of stellar motions on the proplyd exposure time was first
explored in detail by St\"orzer \& Hollenbach (1999) who noted that
beyond a critical distance, \(\rcrit\), from \thC, the FUV flux is too
weak to heat the disk surface sufficiently for it to escape from the
gravitational potential of the central star (Johnstone \etal\ 1998).
In such a case, the disk evaporation is controlled by the ionizing EUV
flux and the mass loss rate falls with distance as \(r^{-1}\).
St\"orzer \& Hollenbach argue that a proplyd on an eccentric orbit
that moves from larger to smaller radii in the cluster is likely to
have an evaporation time, \(\tevap\), roughly equal to its dynamic
time, \(\tdyn \equiv r/u\), during most of its evolution outside of
\(\rcrit\).  Inside \(\rcrit\) the mass loss is controlled by the FUV
flux and is hence only weakly dependent on distance.  They claim that
this leads to a ``freezing'' of the proplyd for \(r<\rcrit\) since
\(\tdyn\) becomes smaller than \(\tevap\), resulting in the proplyd
disk mantaining a roughly constant size during its crossing of the
cluster core.  They calculate a value \(\rcrit \simeq 0.2\) parsec in
Orion (implying a duration of \(\simeq 8\times 10^4\) years for the
passage from \(\rcrit\) to the cluster center) and proceed to estimate
the disk masses for a sample of 10 proplyds for which they have
determined model mass loss rates (see Section~\ref{sec:Comp-with-prev}
above), obtaining values in the range \(0.002\mbox{--}0.01 M_\odot\).
Such a procedure is the converse of the approach adopted here, in
which we proceed from the mass loss rates and the mm-continuum disk
masses in order to estimate the exposure time to ionizing radiation. 

The disk masses obtained by St\"orzer \& Hollenbach (1999) for their
sample seem to be consistent with the estimates from mm-continuum
observations (Section~\ref{sec:Disk-masses}).  However, there are two
problems with their analysis, which may result in their mass loss rate
being underestimated by a factor of about 8.  Firstly, rather than use
the current mass loss rate in their calculations, they use the mass
loss rate that the proplyd would have when placed at a distance
\(\rcrit\), {\em but using the current value for the disk radius\/}.
This ignores the fact that most of the proplyds in their sample are at
\(r \ll \rcrit\) so that, even though \(\tevap > \tdyn \) inside of
\(\rcrit\), these proplyds will have lost half their mass and shrunk
in radius by a factor of 4 since they were at \(\rcrit\).  Secondly,
their mass loss rates are generally smaller than the rates that we
determine for the same proplyds by a factor of at least 2.  As
discussed in Section~\ref{sec:Comp-with-prev}, this is probably mainly 
due to their neglect of the flow from the back side of the disk.
Taking these two factors into account results in disk masses that are
now in conflict with those estimated from mm-continuum observations.
This is consistent with the fact that the \(\simeq 8\times10^4\) year
crossing time for the cluster core is large compared to the \(\simeq
2\times 10^4\) year disk evaporation time calculated in
Section~\ref{sec:Disk-masses}. 
 
It is notable that for all proplyds in our spectroscopic sample except
170--337 the proplyd \OI\ line is significantly blueshifted by 3--7
\kms\ with respect to the nebula (Figure~\ref{fig:line-parameters}).
The nebular \OI\ emission (Wen \& O'Dell 1992) is believed to
originate from the PDR behind the IF on the surface of the background
molecular cloud OMC--1 and is typically blueshifted by 1--2 \kms\ with
respect to the CO emission (Sugitani \etal\ 1986; O'Dell \etal\ 
1993a). If the [\ion{O}{1}] emission from the proplyds traces the
velocity of the enclosed low-mass star, then this implies that 3 of
the 4 proplyd stars in our sample are moving away from the molecular
cloud at 5--9 \kms. These velocities are rather large compared with
the velocity dispersion implied by proper motion studies (see above).
Re-analysis of published proper motion data (Jones \& Walker 1988) for
a sample of proplyds (\(N=21\)) and a sample of non-proplyd stars
(\(N=35\)) lying within the core of the nebula shows no significant
difference between the velocity dispersion of the two samples
(\(2.8\pm 0.5 \)\kms\ for the proplyds; \(2.7\pm 0.3\)\kms\ for the
non-proplyds).  Although many proplyds do have quoted proper motions
larger than this, they also have large estimated uncertainties in the
proper motion (following Jones \& Walker, only stars with
uncertainties \(< 0.1\arcsec/\mbox{century}\) are included in our two
samples). It seems likely then that 3 of our 4 proplyds are very
atypical in their high velocities away from the background molecular
cloud. This need not be surprising since these 3 objects are among the
largest proplyds and hence may have only recently been exposed to
ionizing radiation. One would therefore expect these proplyds to
either lie close to the background IF or to have high blue-shifted
velocities (away from the molecular cloud), or both.

Such an interpretation of the \OIshort\ velocity shift in terms of the
stellar motion depends on there being no systematic asymmetry in the
line profile due to the kinematics of the neutral outflow. Such an
asymmetry could arise, for instance, if the \OIshort\ emission came
predominantly from the base of the neutral flow. In this case, the
receding portion of the flow would be occulted by the opaque disk,
leading to a systematic blueshift in the line with respect to the
stellar velocity. However, the observations of Bally \etal\ (1998)
show that the majority of the \OI\ emission comes from the IF and that
region of the PDR just behind it, rather than from material near the
disk.  In this case, the upper limits established in
sections~\ref{sec:extinct} and \ref{sec:Model-fits} on the optical
depth through the neutral envelope imply that the extinction-induced
asymmetry in the line should be small, which is consistent with the
appearance of the \OI\ line.  Although this line does show a slightly
enhanced blue wing, the difference between the mean velocity and the
peak velocity of the line is in all cases less than 2\kms.  If the
peak \OIshort\ velocity were a truer tracer of the stellar velocity
than the mean \OIshort\ velocity, then this would reduce the
blueshifts of the stars with respect to the molecular cloud to an
average of \(5 \kms\) for the 3 high-velocity proplyds, leading to an
exposure time of approximately \(6000 (z/10^{17} \mbox{cm}) \) years,
where \(z\) is the current distance of the proplyd from the background
IF. This could be reconciled with the exposure time estimated in
Section~\ref{sec:Disk-masses} if \(z \le 2\times 10^{17}\mbox{cm}\),
which would place all 3 proplyds behind \thC, which lies at \(z \simeq
3\mbox{--}6\times 10^{17}\mbox{cm}\).  The inclination angle of
\(75^\circ\) determined for 177--341 (Section~\ref{sec:Model-fits}) is
consistent with this if the distance between \thC\ and the background
IF is at the low end of the estimated range.

\subsubsection{Is there still a ``lifetime problem''?}
\label{sec:still-prob}

The previous two sections indicate that none of the mechanisms that
have been proposed to reduce the time for which the proplyds have been
exposed to ionizing radiation are capable of producing exposure times
as short as the one or two times \(10^4\) years that are demanded by
our measured mass loss rates, coupled with the masses determined from
mm-continuum observations.  Hence the ``lifetime problem'' has not
gone away, and the undeniable existence of the proplyds leaves one in
the situation of having to reject either the mass loss rates or the
measured masses.  The numerical factors involved are not large: a
reduction of one or the other by a factor of 5 would be sufficient to
render plausible either the ``young \ion{H}{2} region'' or the
``moving stars'' hypothesis.  We maintain, however, that the uncertainty
in our mass loss determinations are much less than this, depending as
they do only on the size of the IF and the density and velocity of the
ionized gas, which have all now been directly measured. 

For well-resolved proplyds, the only significant error in the IF size
is due to the assumed distance to the nebula.  Since we adopt a value
of 430 pc, which is at the low end of published estimates, any error
here is only likely to be in the direction of our having {\em
  underestimated\/} the mass loss rate.  We have determined the
ionized density in two different ways. The principal method is to
determine the emission measure of the proplyd cusp from the \Ha\ 
surface brightness, after correcting for foreground extinction.  This
has uncertainties of order 20\% due to the effects of geometry and the
possibility of extinction local to the ionized flow (Henney \& Arthur
1998).  However, since we employed the dust-free value in our
calculations, we are again more likely to have underestimated rather
than overestimated the mass loss rate.  Furthermore, the density
so-determined is consistent with an independent determination from the
C\(^{++}\) doublet ratio in 159--350
(Section~\ref{sec:Valid-dens-deriv}).  The velocity of the ionized
flow just outside the IF is constrained to be \(13 \mbox{--} 16 \kms
\) from our spectroscopic measurements
(Sections~\ref{sec:Extr-Profiles} and \ref{sec:Model-fits}).  Again,
we use the lower value when calculating the mass loss rates.  Although
the outflow velocity has only been measured directly in four objects,
which are atypically large, there is no reason to expect that this
velocity will be any different in the smaller proplyds.  It is
possible that we have overestimated the mass loss rate through the
proplyd tails, since the observed tails show a tapering, which is not
present in the models and may result in the area of the IF in the tail
being less than we have supposed.  However, since the tails only
amount for about half the proplyd mass loss, reducing the tail area by
even a factor of 2 would only diminish the total mass loss rate by
25\%.

In summary, the uncertainty in our mass loss rate determinations for
individual proplyds is probably less than 50\% and the rates are more
likely to be underestimated than overestimated.  There is no apparent
potential source of error that could have led to our systematically
overestimating the mass loss rates by a significant amount.  We
conclude, therefore, that any error must lie in the disk mass
estimates.  Indeed, even disregarding uncertainties in the disk
temperature and the importance of optical depth effects (see
Section~\ref{sec:Disk-masses}), these estimates are completely
dependent on the assumed value of the dust opacity at mm wavelengths.
This opacity could quite easily be a factor of 5 smaller than the
canonical value of \(0.02\,\mbox{cm}^2\mbox{g}^{-1}\) (Beckwith 1999),
either because of a reduced dust-gas ratio or because of the dust
optical properties differing from those supposed.

\subsection{Future work}
\label{sec:Future-work}

Clearly, there are many unanswered questions concerning the mass loss
rate and lifetime of the Orion Nebula proplyds.  We can identify the
following as approaches that will help resolve the situation. We need
accurate emission line profiles for more proplyds than 177--341 and
this expanded sample needs to include a variety of intrinsic distances
from \thC. The key to obtaining this improved sample is through more
accurate subtraction of the nebular background.  This can best be done
using the HST and we are able to report that Cycle 8 time has been
allocated for just that, but only for the object 167--317, which lies
close to \thC. It will be observed using the density sensitive doublet
C~III] at 1909\AA. The power of the Keck I+HIRES system would also
allow measurement of more proplyds from the ground. We need to
determine the true radial velocity of \thC, which is presently highly
uncertain, the most recent determinations being very different from
earlier values (Stahl, \etal\ 1996, O'Dell 1999), with both the old
and new values indicating large relative motions (but in opposite
directions) with respect to the molecular cloud. We also need to have
a self consistent model of the main body of the nebula, i.e. one that
can explain both the current structure and flow of material, which
will also demand explaining the small scale line broadening that may
arise from Alfv\'en waves. It is important, too, to pursue
observations and models that can guide us in understanding the role of
extinction within the proplyd material. Finally, more reliable
estimates of the disk masses are vital, although obtaining these poses 
formidable observational and theoretical challenges.

\acknowledgments WJH acknowledges financial support from DGAPA-UNAM
project IN128698 and from CONACyT project E-25470, Mexico; CRO's work
was supported in part by NASA grant NAG5-1626 and STScI grant
GO-06603.02 to Rice University.  We are very grateful to Jane Arthur,
Vladimir Escalante, David Hollenbach, Doug Johnstone, and Herbert
St\"orzer for many useful suggestions and discussions.  Brian Cudnik
assisted in the early steps of data reduction and Francisco Valdes of
NOAO prepared the special procedures for reducing the two dimensional
HIRES spectra.  We are indebted to the generosity of both John
Stauffer of the Center for Astrophysics for his cooperation in shared
night observing with the Keck and Robert Rubin for allowing us to use
his unpublished STIS spectra of 159--350.

\newpage

\IfPP{
  \newlength{\FigWidth}
  \renewcommand{\figcaption}[2][FileNotFound.ps]{\begin{figure}
      \includegraphics[width=\FigWidth]{#1}
      \caption{#2}
    \end{figure}}
  \nonstopmode 
  }

\IfPP{\setlength{\FigWidth}{\textwidth}}
\figcaption[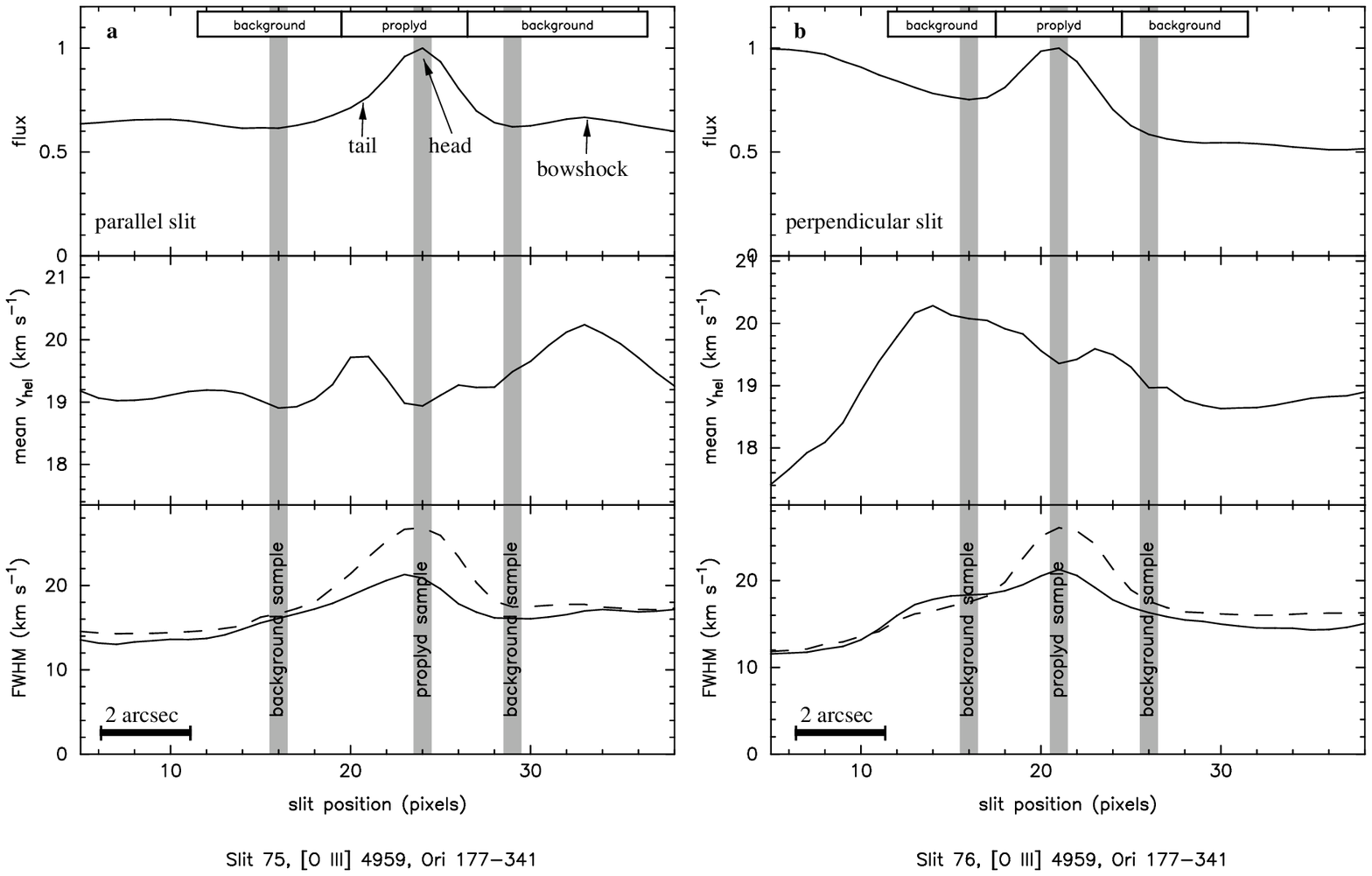]{(a) Line profile parameters as a function of
  position along the spectrograph slit for an example spectrum: \OIIIp\
  from Ori 177--341, with slit oriented parallel to the proplyd axis.
  The samples used in the ``large sample'' method are indicated by the
  white boxes in the top panel, while those used in the ``small
  sample'' method are indicated by the gray strips.  The parameters
  shown are total line intensity (top panel), flux-weighted mean line
  velocity (center panel), and line FWHM (bottom panel, solid line).
  Also shown in the bottom panel (dashed line) is the quantity
  2.355$\sigma$, where $\sigma$ is the flux-weighted RMS velocity 
  width of the line.  This quantity is equal to the FWHM for a Gaussian
  line.  (b) Same as (a), but with slit oriented perpendicular to the 
  proplyd axis. \label{fig:example-samps}}

\IfPP{\setlength{\FigWidth}{\textwidth}}
\figcaption[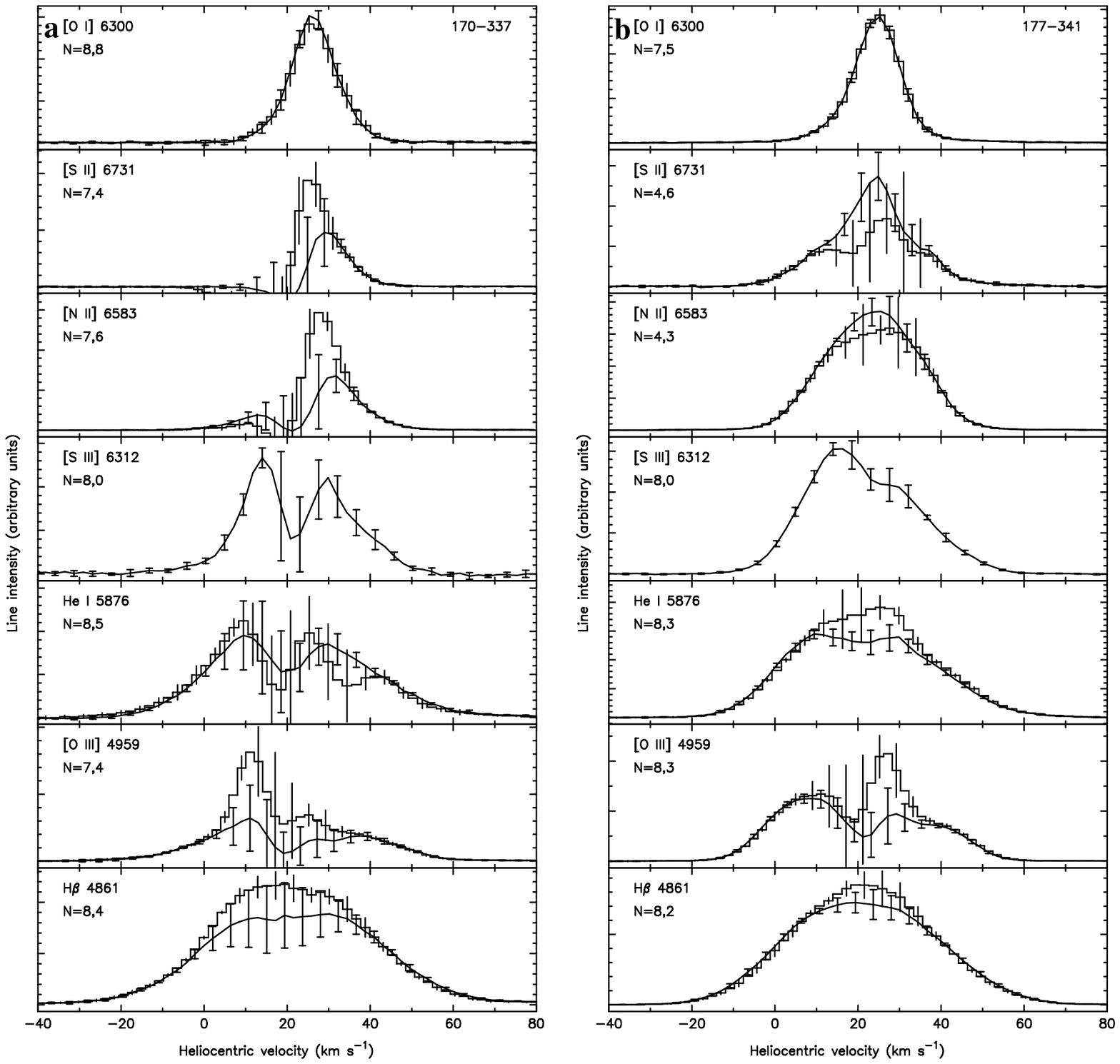]{Proplyd emission line profiles after
  subtraction of nebular background emission for (a) 170--337, (b)
  177--341, (c) 182--413, and (d) 244--440. Results of the
  ``small-sample'' subtraction method are shown by continuous lines
  and error bars with terminators.  Results of the ``large-sample''
  subtraction method are shown by stepped histograms and error bars
  without terminators.  The error bars represent the RMS deviation
  among individual samples, with the number of samples used for the
  two methods being indicated for each profile.  The lines are
  arranged in order of degree of ionization of the emitting ion, with
  the exception of the Balmer lines (\Ha, or \Hb\ if \Ha\ was
  saturated), which are placed last due to their large thermal widths.
  The number of independent samples used by, respectively, the
  ``small-sample'' and ``large-sample'' methods are also indicated on
  each spectrum (note that only the ``small-sample'' method was used
  for \SIIIp). \label{fig:line-profiles} }

\addtocounter{figure}{-1}
\figcaption[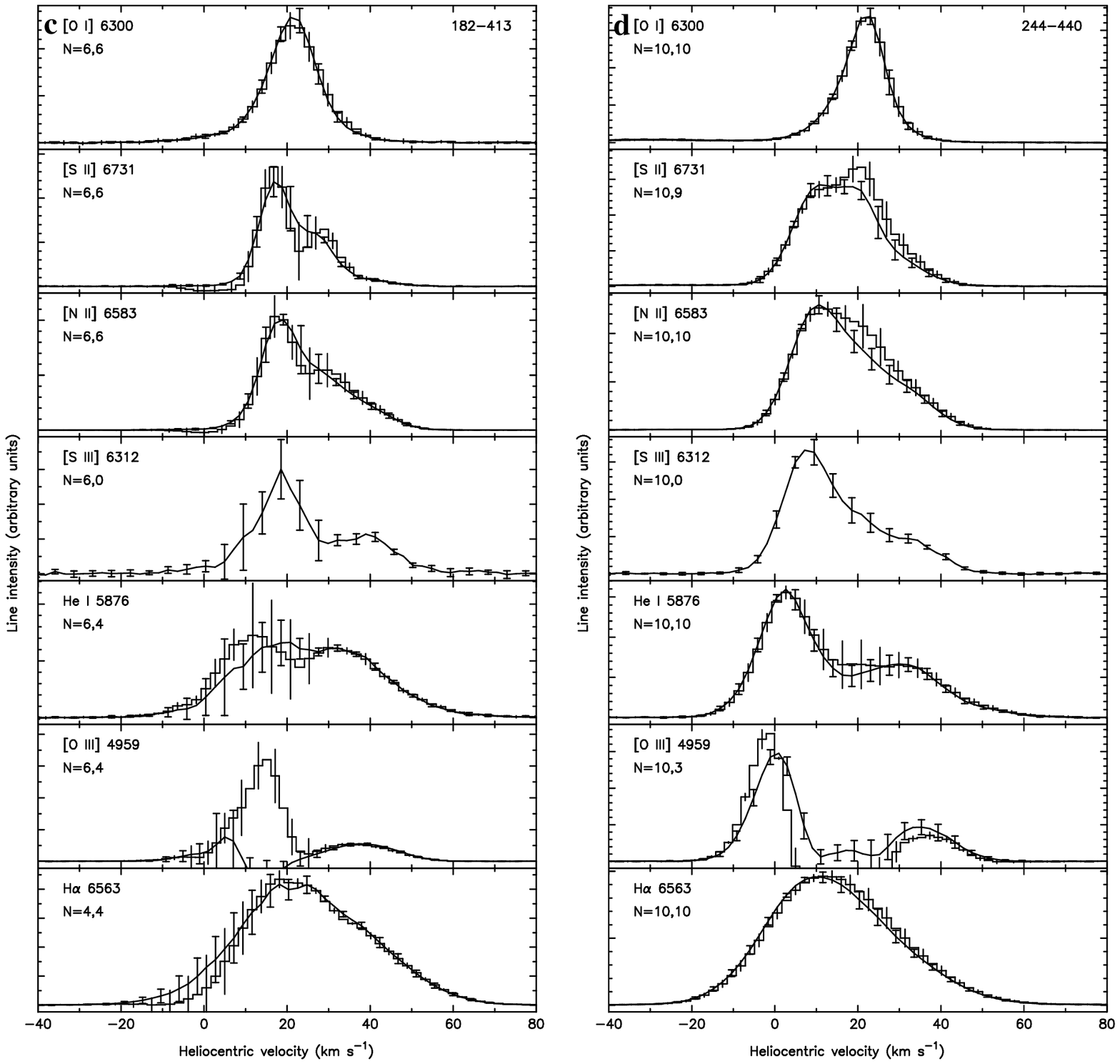]{(cont.)}

\IfPP{\setlength{\FigWidth}{0.95\textwidth}}
\figcaption[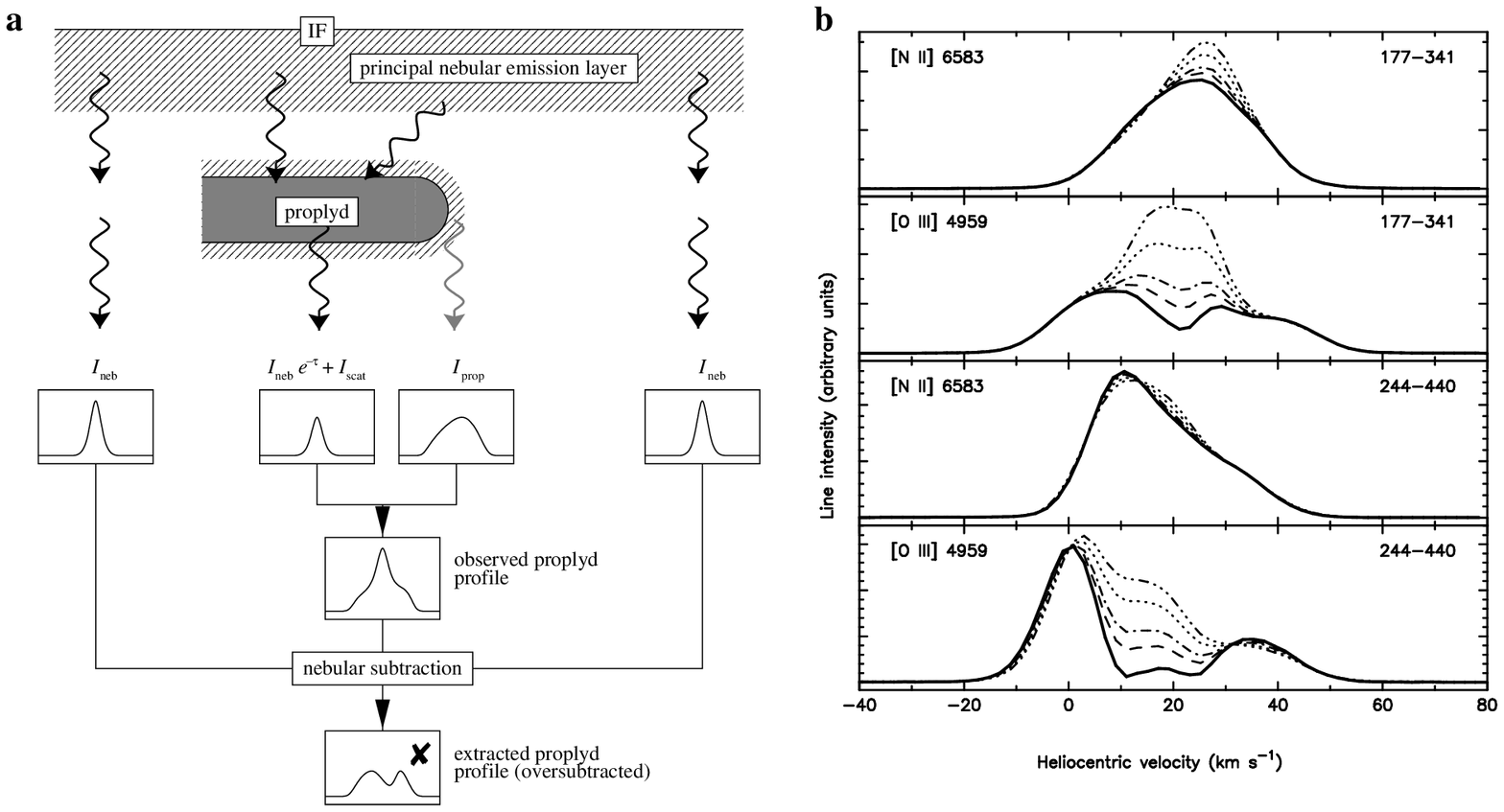]{Effects on the extracted line profiles of
  the assumed extinction of the background nebula by dust in the
  proplyd.  (a) Schematic illustration of how the neglect of internal
  extinction in the proplyd can lead to the oversubtraction of the
  nebular emission component if the principal nebular emission layer
  lies behind the proplyd.  (b) Results of assuming different values
  for the effective extinction in the proplyd for the \NIIp\ and
  \OIIIp\ lines of 177--341 and 244--440.  For each panel, the thick
  line corresponds to the no-extinction case and the thin lines
  correspond to effective visual extinctions through the proplyd of
  \(\Av =\) 0.1 (dashed), 0.2 (dot-dashed) , 0.5 (dotted), 1.0
  (dot-dot-dot-dashed). For increased assumed extinction, less nebular
  background is subtracted from the observed profiles, leading to more
  emission being attributed to the proplyd.
  \label{fig:dust}}

\IfPP{\setlength{\FigWidth}{0.95\textwidth}}
\figcaption[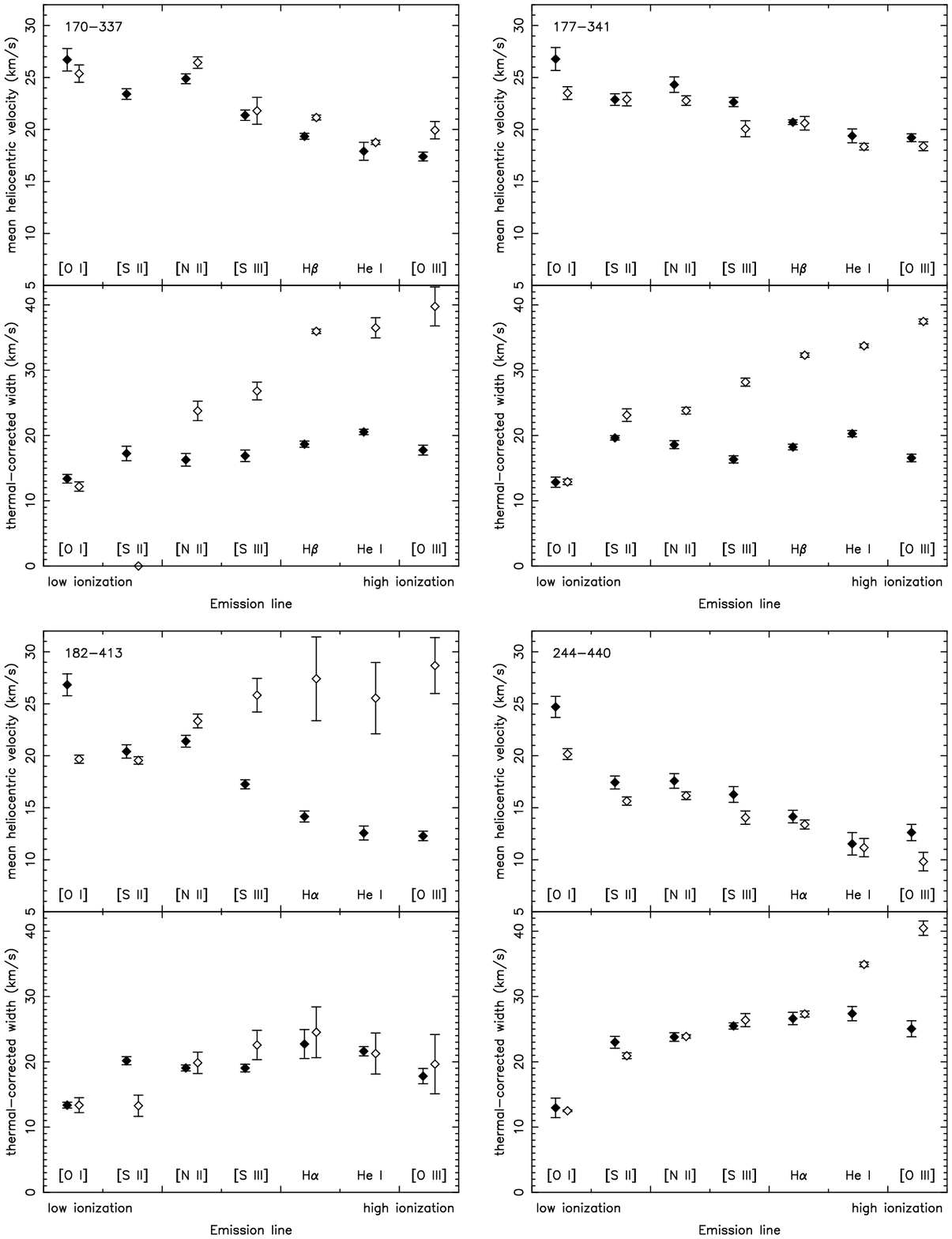]{Mean velocity and velocity width as a
  function of degree of ionization for the extracted proplyd spectra
  (open symbols) and adjacent nebula emission (filled symbols). (a)
  170--337, (b) 177--341, (c) 182--413, and (d) 244--440. The velocity
  width has been corrected for instrumental and thermal broadening
  (assuming $T = 8900$K).  The results for \HeIp\ have also been
  corrected for the effects of fine-structure splitting.  Error bars
  merely show the RMS variation among individual samples and take no
  account of possible systematic errors.
  \label{fig:line-parameters}}

\IfPP{\setlength{\FigWidth}{\textwidth}}
\figcaption[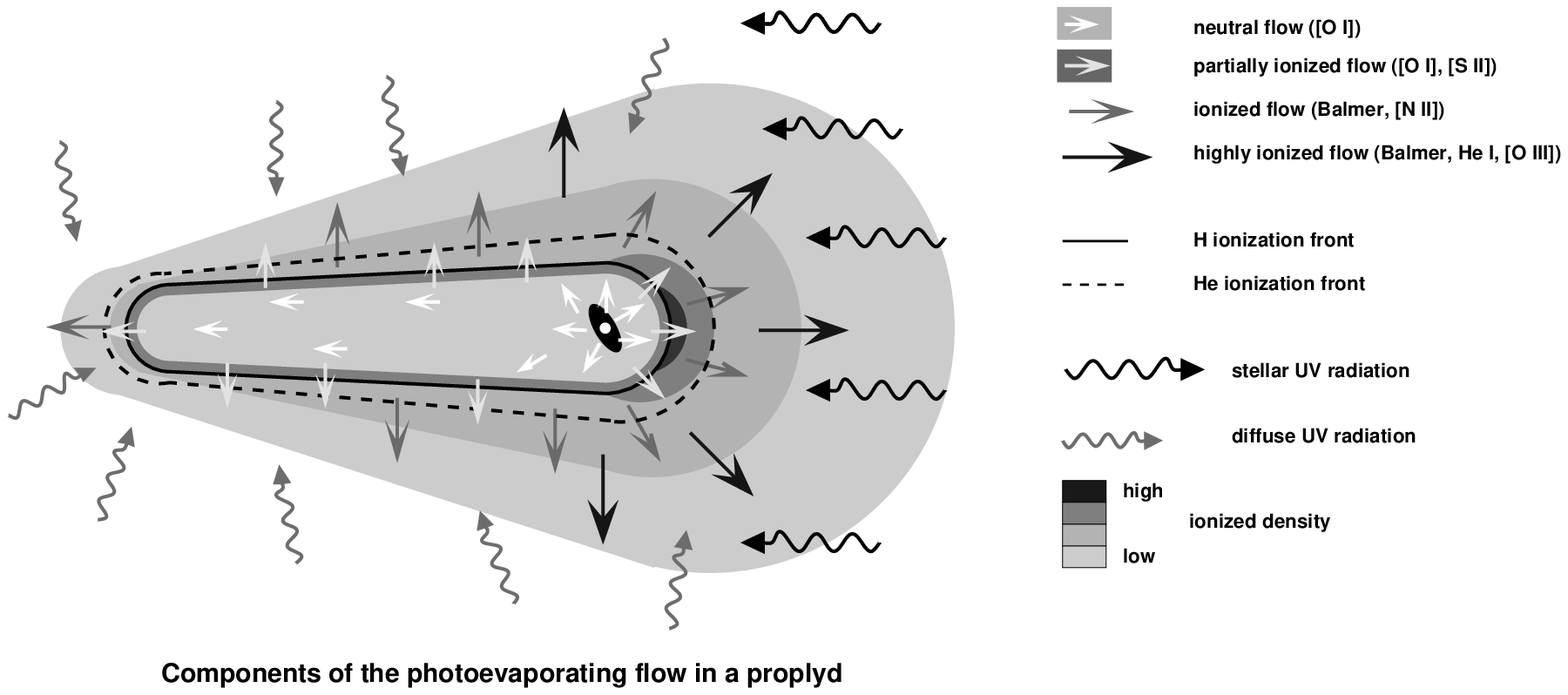]{Schematic representation of the
  photoevaporating flow model, showing the principal flow components.
  Stellar EUV and FUV photons enter from the right. The FUV photons
  penetrate to the surface of the accretion disk around the low-mass
  star (Johnstone \etal\ 1998), driving a slow neutral flow, which,
  for most proplyds (Henney \& Arthur 1997; St\"orzer \& Hollenbach
  1999), accelerates to mildly supersonic velocities before shocking
  and passing through an ionization front (IF) at a distance of a few
  disk radii. In the IF the gas is rapidly accelerated to about 13
  \kms, and continues to accelerate as it expands away from the IF and
  reaches progressively higher stages of ionization. The neutral flow
  in the tail is fed by diffuse UV photons, which evaporate the back
  side of the disk, and possibly also by gas that left the front side
  of the disk but was redirected into the tail by pressure gradients
  in the shocked neutral layer (Bertoldi \& Johnstone 1999). The
  ionized flow from the tail is induced by diffuse EUV photons, but
  stellar EUV photons entering from the side also play an important
  role in mantaining the ionization of the tail flow once it has left
  the IF, especially towards the front of the tail (Henney 1999).
  \label{fig:cartoon}}

\IfPP{\setlength{\FigWidth}{\textwidth}}
\figcaption[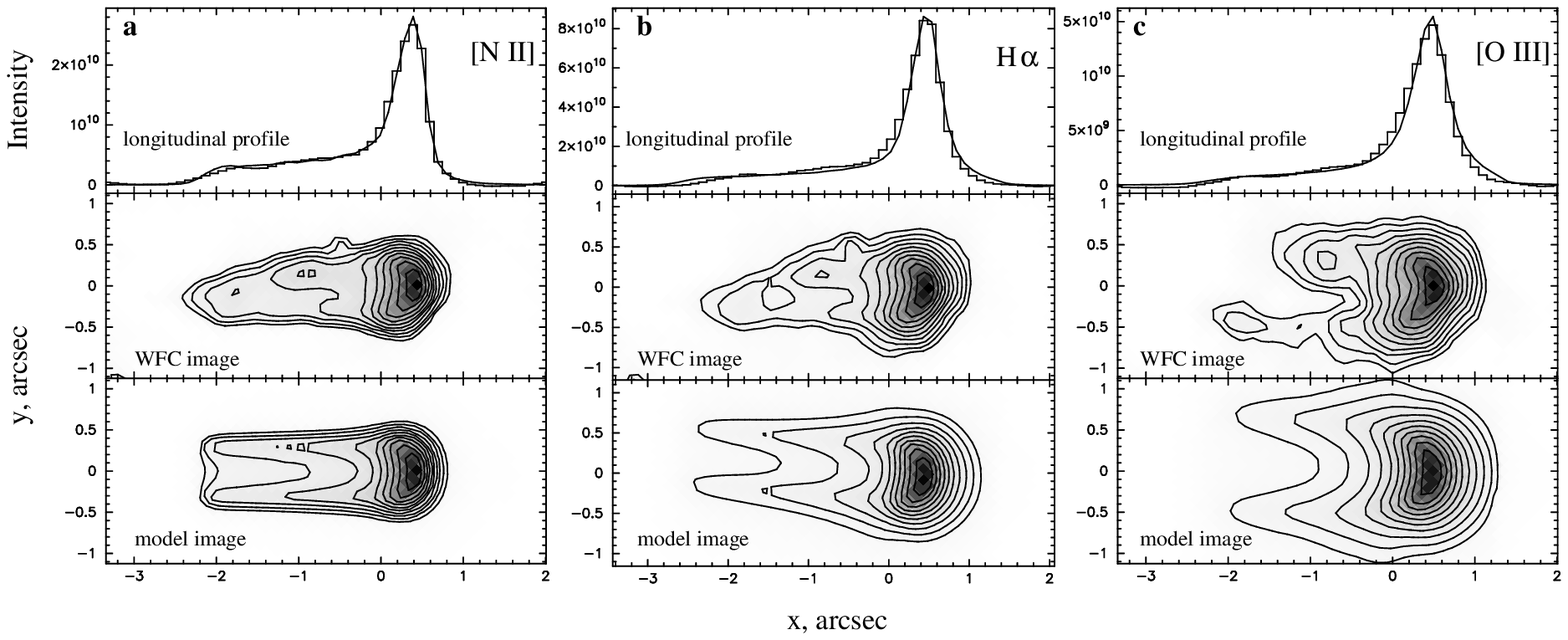]{Comparison between photoevaporating flow
  models and flux-calibrated, background-subtracted \HST\ WFC images
  of the proplyd 177--341. (a) \NIIp, (b) \Ha, and (c) \OIIIp.  For
  each emission line, the top panel shows the intensity profile (phots
  s$^{-1}$ cm$^{-2}$ sr$^{-1}$) along the long axis of the proplyd,
  with the observations shown by the stepped histogram and the model
  fit by the continuous line.  The center and bottom panels
  respectively show the observed and model image, with linear
  grayscale and logarithmic contours both representing the image
  intensity.  The interval between adjacent contours is $2^{1/2}$ and
  the lowest contour is $1/32$ of the peak intensity.  The nebular
  background subtraction and automated model fitting were performed as
  described in Henney \& Arthur (1998).  The models shown have
  $i=75^\circ$ and the other parameters have the best-fit values
  listed in Table~\protect\ref{tab:HST}.\label{fig:WFC}}

\IfPP{\setlength{\FigWidth}{\textwidth}}
\figcaption[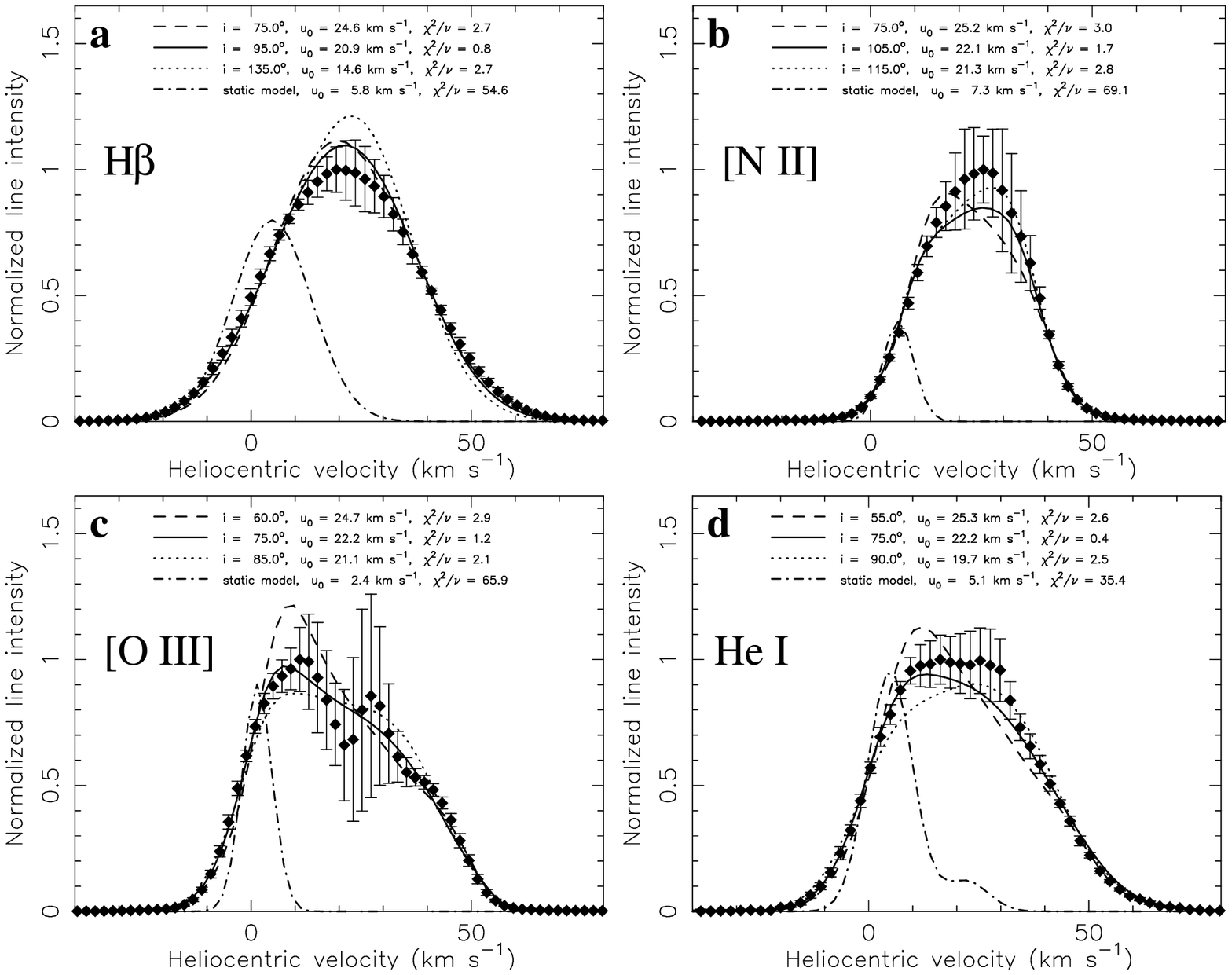]{Model fits to extracted line profiles for
  the proplyd 177--341. (a) \Hb\ (b) \NIIp\ (c) \OIIIp\ (d) \HeIp. The
  observed extracted line profiles (see
  Section~\ref{sec:Extr-Profiles}) are shown by solid symbols. The
  model profile for the best-fit value of $i$ is shown by the solid
  line, while the dashed and dotted lines show respectively the
  smallest and largest values of $i$ with reduced \(\chi^2 < 3\). The
  dot-dashed line shows the best-fitting static model.  The observed
  profiles assume a mean extinction of the background by dust in the
  proplyd of \(A_V = 0.1\).  The models have $M_0 = 1$ and the other
  parameters have the best-fit values listed in
  Table~\protect\ref{tab:HST}.
  \label{fig:fit177}}

\IfPP{\setlength{\FigWidth}{0.5\textwidth}}
\figcaption[Henney.fig8.ps]{Histogram of proplyd mass loss rates from our
  spectroscopic sample ($N=4$) and the Henney \& Arthur (1998) sample
  ($N=27$, excluding 2 objects that overlap with the spectroscopic
  sample).  In all cases, the mass loss rates were calculated by
  scaling the result for 177--341 according to the measured value of
  $n_0 r_0^2 $ for each proplyd.  For those objects that lack
  prominent tails ($N=5$), the scaled value was divided by 2.  Taking
  into account the uncertainties in $n_0$, $r_0$, and $u_0$, the the
  uncertainty in the mass loss rate for individual sources is about
  $\pm 50\%$, or $\pm 1$ bin in the histogram. The mean mass loss rate 
  is $3.2\times 10^{-7} \smy$ for the Henney \& Arthur sample and
  $9.8\times 10^{-7} \smy$ for the spectroscopic
  sample. \label{fig:mass-loss}} 

\end{document}